\documentclass[12pt,a4paper]{article}

\usepackage{amsmath,amssymb,amsthm}
\usepackage{bm}
\usepackage[dvips]{graphicx}
\usepackage{wrapfig}
\usepackage{fancybox}
\usepackage{ascmac}
\usepackage{framed}
\usepackage{float}
\usepackage[mathscr]{euscript}
\usepackage{amscd}
\usepackage{mathrsfs}
\usepackage{pifont}
\usepackage{braket}
\usepackage[usenames]{color}

\allowdisplaybreaks[4]

\makeatletter
\def\preprint#1{\def\@preprint{#1}}
\def\department#1{\def\@department{#1}}

\def\@maketitle{ 
\begin{flushright} 
{\@preprint}  
\end{flushright}
\vspace{5mm}
\begin{center} 
{\LARGE \@title \par}
\vspace{7mm}
{\large \@author \par}
\vspace{7mm}
{\large \@department \par}
\vspace{7mm}
{\large \@date}
\end{center}
 }
\makeatother

\preprint{
KOBE-TH-13-02\\
KUNS-2435\\
HRI-P-13-02-001\\
RECAPP-HRI-2013-002\\}

\title{\bf Shifted orbifold models with magnetic flux}

\author{
Yukihiro FUJIMOTO${}^a$\footnote{E-mail:093s121s@stu.kobe-u.ac.jp}, 
Tatsuo KOBAYASHI${}^b$\footnote{E-mail:kobayash@gauge.scphys.kyoto-u.ac.jp}, 
Takashi MIURA${}^a$\footnote{E-mail:takashi.miura@people.kobe-u.ac.jp},\\ 
Kenji NISHIWAKI${}^c$\footnote{E-mail:nishiwaki@hri.res.in} 
~and Makoto SAKAMOTO${}^a$\footnote{E-mail:dragon@kobe-u.ac.jp}}

\department{
{\it ${}^a$Department of Physics, Kobe University, Kobe 657-8501, Japan}\\
{\it ${}^b$Department of Physics, Kyoto University, Kyoto 606-8502, Japan}\\
{\it ${}^c$Regional Centre for Accelerator-based Particle Physics, }\\
{\it Harish-Chandra Research Institute, Allahabad 211 019, India}\\}

\date{February 22, 2013}
\makeatletter
\@addtoreset{equation}{section}
\@addtoreset{figure}{section}
\@addtoreset{table}{section}
\makeatother

\begin{document}

\maketitle

\vspace{10mm}
\begin{center}
Abstract
\end{center}
{\rm We propose a mechanism to obtain the generation of matter in the standard model.
We start from the analysis of the $T^2/Z_N$ shifted orbifold with magnetic flux, which imposes a $Z_N$ symmetry on torus.
We also consider several orbifolds such as $(T^2\times T^2)/Z_N$, $(T^2 \times T^2
\times T^2)/(Z_N \times Z_{N'})$ and $(T^2 \times T^2 \times T^2)/(Z_N
\times Z_{N'} \times Z_{N''})$.
On such orbifolds, we study the behavior of fermions in two different means, one is the operator formalism and the other is to analyze wave functions explicitly.
For an interesting result, it is found that the number of zero-mode fermions is related to $N$ of the $Z_N$ symmetry.
In other words, the generation of matter relates to the type of orbifolds.
Moreover, we find that shifted orbifold models to realize three generations are, in general, severely restricted. 
For example, the three-generation model on the type of $M^4 \times (T^2 \times T^2)/Z_N$ is unique.
One can also construct other types of three-generation orbifold models with rich flavor structure. 
Those results may bring us a realistic model with desired Yukawa structure.}

%Similarly one can construct the models with the same number of
%generations 
%through other orbifolds and the flavor structure becomes rich.
%These results may bring us one of a few possibilities for realizing of the three-generation model.}

\pagebreak
%%%%%%%%%%%%%%%%%%%%%%%%%%%%%%%%%%%%%%%%%%%%%%%%%%%%%%%%%%%%%%%%%%%%%%%%%%%%%%%
\section{Introduction}

Extra dimensional field theories play an important role in 
particle physics and cosmology.
In particular, string-derived higher dimensional models 
would be interesting.

It is a key issue how to realize a 4D chiral theory 
through a certain type of compactification, 
when we start with a higher dimensional theory.
For example, the toroidal compactionfication is 
simple, but it does not lead to a 4D chiral theory 
without any gauge background.
More complicated geometrical backgrounds such as 
Calabi-Yau compactifications can lead to 
a 4D chiral theory.
In general, it is difficult to solve 
zero-mode equations analytically on Calabi-Yau manifolds, 
although topological discussions are applicable.

The torus compactification with magnetic fluxes 
is of quite interesting backgrounds as extra dimensional 
models \cite{Cremades:2004wa}.
Chiral massless spectra in a 4D theory can be realized.
One can solve zero-mode equations analytically 
and their zero-mode profiles are non-trivially quasi-localized.
Furthermore, such supersymmetric Yang-Mills models 
on the torus with magnetic fluxes can be derived from 
the superstring theory with D-branes \cite{Blumenhagen:2000wh}.
Also, the general form of wave functions and the spectra 
have been derived in arbitrary $n$-dimensional torus 
with magnetic flux \cite{Sakamoto:2003}.

The number of zero modes is determined by the magnitude 
of the magnetic flux.
Thus, one can realize the three generations of chiral matter fields 
by choosing a proper pattern of magnetic fluxes.\footnote{
On the other hand, there have been some bottom-up approaches 
realizing the three generations of chiral matter fields, 
what is called the extra dimension models.
For example, an attempt is to derive the three generations 
of chiral matter fields on the basis of the relation between gauge symmetry 
and boundary conditions on four dimensional spacetime 
added to $S^1/Z_2$ twisted orbifold \cite{Kawamura:2007}.   
Another one is to derive the three generations 
of chiral matter fields by effect of geometry of extra dimension 
with point interactions instead of magnetic flux
\cite{Fujimoto:2012}.}
In addition, non-trivial profiles of quasi-localized zero-modes 
can lead to hierarchically small couplings when 
they are localized far away from each other, although 
their couplings would be of ${\cal O}(1)$ for zero-modes localized 
near places.
Thus, these models are phenomenologically quite interesting.
Indeed, several studies have been 
carried out for various phenomenological aspects 
such as explicit model building and computations 
of 4D low-energy effective field theories, e.g. 
Yukawa couplings \cite{Cremades:2004wa}, 
realization of quark/lepton masses 
and their mixing angles \cite{Abe:2012fj}, 
higher order couplings \cite{Abe:2009dr}, 
flavor symmetries \cite{Abe:2009vi,Abe:2009uz,BerasaluceGonzalez:2012vb}, 
massive modes \cite{Hamada:2012wj}, etc.
(See also \cite{Antoniadis:2004pp,Choi:2009pv,DiVecchia:2011mf,Abe:2012ya}.)

The orbifold compactifiction with magnetic flux is 
also interesting \cite{Abe:2008fi,Abe:2008sx}.
The (twisted) orbifold is constructed by dividing the 
torus by a discrete rotation  \cite{Choi:2006qh}.
On the $Z_2$ orbifold with magnetic flux, 
zero-mode wave functions are classified into 
even and odd functions under the $Z_2$ reflection.
Either $Z_2$ even or odd zero modes are projected out 
exclusively by the orbifold projection.
Then, we can obtain that the number of generations differs from 
one in the simple toroidal compactification with the same 
magnetic flux.
Hence, the flavor structure becomes rich.

In this paper, we study another type of orbifolds 
with magnetic flux.
Instead of the discrete rotation, we divide 
the simple toroidal compactification with magnetic flux 
by some discrete shift symmetries such as $Z_N$, 
i.e. the shifted orbifold.\footnote{
For example, the shifted orbifolds have been studied 
within the framework of the heterotic string theory 
\cite{Schellekens:1987ij,Chun:1990iw, Kobayashi:1991yg}, 
and also in the context of the torus-orbifold equivalence \cite{ Kobayashi:1987dx, Imamura:1991}.}
We can classify zero-mode wave functions by their behaviors 
under the shift symmetry and project out some of them.
Then, we derive a new type of models and flavor structures.
In particular, we consider $T^2/Z_N$, $(T^2\times T^2)/Z_N$, $(T^2\times T^2)/Z_N\times T^2$, 
$(T^2 \times T^2 \times T^2)/(Z_N \times Z_{N'})$ and 
$(T^2 \times T^2 \times T^2)/(Z_N \times Z_{N'} \times Z_{N''})$ shifted orbifolds.
Then, we study the number of zero modes and their 
wave functions on the above shifted orbifolds with magnetic fluxes.
For an interesting result, it is found that the number of zero-mode 
fermions is related to $N$ of the $Z_N$ symmetry.
In other words, the generation of matter relates to the type of orbifolds.
Moreover, we find that shifted orbifold models to realize three generations are, in general, severely restricted. 
For example, the three-generation model on the type of $M^4 \times (T^2 \times T^2)/Z_N$ is unique.
One can also construct other types of three-generation orbifold models with rich flavor structure. 
Those results may bring us a realistic model with desired Yukawa structure.

This paper is organized as follows.
In section 2, we study the shifted orbifold with magnetic fluxes 
in the operator formalism.
In section 3, we study zero-mode wave functions of 
spinor fields and flavor structure.
Section 4 is devoted to conclusion and discussion.
In Appendix A, we discuss the general form of $Z_N$ shifted orbifold and basis transformation.
In Appendix B, we discuss the degeneracy of spectrum on $(T^2 \times T^2 \times T^2)/(Z_N \times Z_{N'} \times Z_{N''})$.
In Appendix C, we discuss Wilson line phases and the redefinition of fields under the existence of magnetic flux.

%%%%%%%%%%%%%%%%%%%%%%%%%%%%%%%%%%%%%%%%%%%%%%%%%%%%%%%%%%%%%%%%%%%%%%%%%%%%%%%
\section{Operator formalism on shifted orbifold with magnetic flux\label{Op_form}}

First of all, we introduce a homogeneous magnetic flux along a $U(1)$ gauge group, which may be a subgroup in some non-Abelian gauge group, and concentrate only on the $U(1)$ gauge theory.
In this section, we consider the quantum mechanical system of the $U(1)$ gauge theory on some shifted orbifold with homogeneous magnetic flux.
The shifted orbifold is defined by orbifolding with some discrete symmetries on torus, as we will see below.
For the analysis, we make use of a technique of operator formalism, in which it is easy to understand the number of degeneracy as compared with the analysis with the wave functions. 
We firstly investigate the quantum mechanical system with a homogeneous magnetic field on $T^2/Z_{N}$, and show that the energy spectrum of this system has degeneracy due to the magnetic flux, which is well-known as Landau levels. 
In this case, however, there is no reason to restrict the number of degeneracy. 
After that, we also investigate the system on $(T^2 \times T^2)/Z_{N}$.
Differently from $T^2/Z_N$, we obtain the remarkable result that the number of degeneracy is restricted to a multiple of $N$.
This result implies that the three-generation structure could be derived from some higher dimensional gauge theory.
We also discuss the extension of this idea to $(T^2 \times T^2\times T^2)/(Z_{N}\times Z_{N'})$ as well as $(T^2 \times T^2\times T^2)/(Z_{N}\times Z_{N'} \times Z_{N''})$ in the latter half of this section.

\subsection{$T^2/Z_{N}$ shifted orbifold\label{OP_T2ZN}}

\subsubsection{Operator formalism on $T^2$}

Let us start from an analysis of a quantum mechanical system in a homogeneous magnetic field on $T^2$. 
We define the vector notation on $T^2$ as $\bm{y}\equiv (y_1,y_2)^\mathrm{T}$.
The Schr\"odinger equation and the Hamiltonian we consider are written as
	%%%%%%%%%%%%%%%%%%%%%%%%%%%%%%%%%%%%%%%%%%%%%%
	\begin{align}
	&H\psi ({\bm y}) =E\psi({\bm y}),~~~~~%\notag \\
    H=\left(-i{\bm \nabla} +q{\bm A}({\bm y})\right)^2,
	\label{Schrodinger}
	\end{align}
	%%%%%%%%%%%%%%%%%%%%%%%%%%%%%%%%%%%%%%%%%%%%%%
where $\bm{\nabla}\equiv (\partial_{y_1}, \partial_{y_2})^\mathrm{T}$, $q$ is a $U(1)$ charge and ${\bm A}({\bm y})$ provides a homogeneous magnetic field on $T^2$ as 
	%%%%%%%%%%%%%%%%%%%%%%%%%%%%%%%%%%%%%%%%%%%%%%
	\begin{align}
	{\bm A}({\bm y})=-\frac{1}{2}\Omega {\bm y}+\frac{2\pi}{q}{\bm a},
    %\label{gaugeA}
	\end{align}
	%%%%%%%%%%%%%%%%%%%%%%%%%%%%%%%%%%%%%%%%%%%%%%
or in the component 
	%%%%%%%%%%%%%%%%%%%%%%%%%%%%%%%%%%%%%%%%%%%%%%
	\begin{align}
	A_i({\bm y})=-\frac{1}{2}\sum_{j=1}^2\Omega_{ij} y_j+\frac{2\pi}{q}a_i,~~~~~(i=1,2).
    %\label{gaugeA}
	\end{align}
	%%%%%%%%%%%%%%%%%%%%%%%%%%%%%%%%%%%%%%%%%%%%%%
Here, $\bm{a}$ is a Wilson line phase which is composed of real constants.
We emphasize that only the antisymmetric part of $\Omega$ is physical and the symmetric part depends on a choice of gauge. We will take a suitable gauge as we will see later.
Introducing a two-dimensional lattice
	%%%%%%%%%%%%%%%%%%%%%%%%%%%%%%%%%%%%%%%%%%%%%%
	\begin{align}
	\Lambda =\Bigl\{  \hspace{0.2em} \sum^{2}_{a=1}n_{a}{\bm u}_{a} \hspace{0.2em}\Bigl|\hspace{0.2em} n_{a}\in \mathbb{Z} \Bigr\},
	\end{align}
	%%%%%%%%%%%%%%%%%%%%%%%%%%%%%%%%%%%%%%%%%%%%%%
we define the two-dimensional torus as $T^2 = \mathbb{R}^2 / \Lambda$.
Here $\bm{u}_1$ and $\bm{u}_2$ are the basis vectors of torus. 
In other words, $T^2$ is defined with the identification
	%%%%%%%%%%%%%%%%%%%%%%%%%%%%%%%%%%%%%%%%%%%%%%
	\begin{align}
	{\bm y}\sim {\bm y}+\sum^{2}_{a=1}n_{a}{\bm u}_{a}. % ,\hspace{1.3em}({}^{\forall}n_{a}\in\mathbb{Z} \hspace{0.2em}). 
    \label{identification}
	\end{align}
	%%%%%%%%%%%%%%%%%%%%%%%%%%%%%%%%%%%%%%%%%%%%%%	
	
For the Schr\"odinger equation to be compatible with this condition, the wave function $\psi ({\bm y})$ has to satisfy the pseudo-periodic boundary conditions
	%%%%%%%%%%%%%%%%%%%%%%%%%%%%%%%%%%%%%%%%%%%%%%
	\begin{align}
	\psi({\bm y}+{\bm u}_{a})= e^{iq{\bm y}^{\mathrm{T}}\Omega {\bm u}_{a}/2}\psi({\bm y})~~~~~~\mathrm{for}~a=1,2.
    \label{pseudoperiodicBC}
	\end{align}
	%%%%%%%%%%%%%%%%%%%%%%%%%%%%%%%%%%%%%%%%%%%%%%
Here we made a constant phase appearing at the right-hand side of this equation absorbed into $\bm{a}$.
Now, we require that the wave function must be single-valued on the torus. The requirement leads to the magnetic flux quantization condition
	%%%%%%%%%%%%%%%%%%%%%%%%%%%%%%%%%%%%%%%%%%%%%%
	\begin{align}
	q{\bm u}_{a}^{\mathrm{T}}B{\bm u}_{b}=2\pi Q_{ab},
	\end{align}
	%%%%%%%%%%%%%%%%%%%%%%%%%%%%%%%%%%%%%%%%%%%%%%
where $B\equiv \frac{1}{2} (\Omega- \Omega^{\mathrm{T}})$, which corresponds to a homogeneous magnetic field, and $Q_{ab}=-Q_{ba}\in\mathbb{Z}$.	

\begin{figure}[t]
	\begin{center}
	\includegraphics[width=60mm]{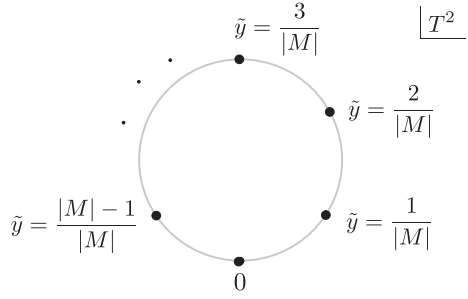}
	\end{center}
	\caption{{\small Schematic figure of the quantization of $\tilde{y}=j/|M|$ .}}
	\label{fig:one}
	\end{figure}

To move on the operator formalism, we introduce a momentum operator as ${\bm p}\equiv -i{\bm \nabla}$, which is  canonically conjugate to ${\bm y}$.
This operator satisfies the canonical commutation relations
	%%%%%%%%%%%%%%%%%%%%%%%%%%%%%%%%%%%%%%%%%%%%%%
	\begin{align}
	[y_{j},p_{k}]&=i\delta_{jk},~~{\rm the~others}=0,~~~~~(j,k=1,2).
	\end{align}
	%%%%%%%%%%%%%%%%%%%%%%%%%%%%%%%%%%%%%%%%%%%%%%
We rewrite the wave function $\psi({\bm y})$ in the language of the operator formalism as $\psi ({\bm y})= \Braket {\bm y|\psi}$ and then the system is rewritten by
	%%%%%%%%%%%%%%%%%%%%%%%%%%%%%%%%%%%%%%%%%%%%%%
	\begin{align}
	&\hat{H}|\psi\rangle =E|\psi\rangle, ~~~~~%\notag \\
	\hat{H}=\left(\hat{{\bm p}}-\frac{q}{2}\Omega \hat{{\bm y}}+2\pi{\bm a}\right)^2\equiv \hat{{\bm p}}'{}^{2}.
	\end{align}
	%%%%%%%%%%%%%%%%%%%%%%%%%%%%%%%%%%%%%%%%%%%%%%
This is restricted by the constraint conditions
	%%%%%%%%%%%%%%%%%%%%%%%%%%%%%%%%%%%%%%%%%%%%%%
	\begin{align}
	&e^{i\hat{T}_{a}}|\psi\rangle =|\psi\rangle, ~~~~~%  \\
	\hat{T}_{a}= {\bm u}^{\mathrm{T}}_{a}\left(\hat{{\bm p}}-\frac{q}{2}\Omega^{\mathrm{T}}\hat{{\bm y}}\right),~~~~~(a=1,2). 
    \label{constraint}%\label{translation}
	\end{align}
	%%%%%%%%%%%%%%%%%%%%%%%%%%%%%%%%%%%%%%%%%%%%%%
These conditions come from the pseudo-periodic boundary conditions (\ref{pseudoperiodicBC}) and we have taken the suitable gauge in which the additional term in $\hat{T}_a$ vanishes, i.e. ${\bm u}^{\mathrm{T}}_{a}\Omega {\bm u}_{a}=0$.
To investigate the eigenstates of this system and their energy eigenvalues, it is convenient to define new variables under a canonical transformation. Then we introduce the new variables
	%%%%%%%%%%%%%%%%%%%%%%%%%%%%%%%%%%%%%%%%%%%%%%
	\begin{align}
	&\hat{Y}\equiv \frac{\sqrt{2}}{\omega }\hat{p}'{}_{2},~~\hat{P}\equiv \sqrt{2}\hat{p}'{}_{1}, \notag \\
	&\hat{\widetilde{Y}}\equiv -\frac{1}{2\pi M}\hat{T}_{2},~~\hat{\widetilde{P}}\equiv \hat{T}_{1},
	\end{align}
	%%%%%%%%%%%%%%%%%%%%%%%%%%%%%%%%%%%%%%%%%%%%%%
where $\omega \equiv 2q B_{12}=2qb/|\bm{u}_1\times \bm{u}_2|$
\footnote{The $b$ is the magnitude of magnetic flux and $b=\int_{T^2}F$ as eq.(\ref{Fb}). }
 and $M\equiv Q_{12}$.
They satisfy the canonical commutation relations, i.e.
	%%%%%%%%%%%%%%%%%%%%%%%%%%%%%%%%%%%%%%%%%%%%%%
	\begin{align}
	[\hat{Y},\hat{P}]=i,~~[\hat{\widetilde{Y}},\hat{\widetilde{P}}]=i,~~{\rm the~others}=0.
	\label{com_T2_YP}
	\end{align}
	%%%%%%%%%%%%%%%%%%%%%%%%%%%%%%%%%%%%%%%%%%%%%%
The transformation $\{y_i,p_i;i=1,2\} \mapsto \{ Y,P,\tilde{Y},\tilde{P}\}$ is a canonical one since it preserves the canonical commutation relations.
Moreover, under the new variables, the system can be reformulated into
effectively one-dimensional harmonic oscillator
	%%%%%%%%%%%%%%%%%%%%%%%%%%%%%%%%%%%%%%%%%%%%%%
	\begin{align}
	&\hat{H}=\frac{1}{2}\hat{P}^2 +\frac{\omega^2}{2}\hat{Y}^2 ,
    \label{harmonicoscillator}
	\end{align}
	%%%%%%%%%%%%%%%%%%%%%%%%%%%%%%%%%%%%%%%%%%%%%%
with the two constraint conditions
	%%%%%%%%%%%%%%%%%%%%%%%%%%%%%%%%%%%%%%%%%%%%%%
	\begin{align}
	&e^{i\hat{\widetilde{P}}}|\psi\rangle=|\psi\rangle,~~
    e^{-i2\pi M\hat{\widetilde{Y}}}|\psi\rangle=|\psi\rangle
    \label{constraint1}.
	\end{align}
	%%%%%%%%%%%%%%%%%%%%%%%%%%%%%%%%%%%%%%%%%%%%%%	
These two constraint conditions are, however,  independent of the energy
spectrum of the harmonic oscillator because $\hat{H}$ is
constructed only from $\hat{Y}$ and $\hat{P}$, but not from $\hat{\widetilde{Y}}$ and $\hat{\widetilde{P}}$. 
The constraint conditions (\ref{constraint1}) lead to the coordinate quantization as below.
We take the coordinate representation $|\tilde{y}\rangle$ which diagonalizes the operator $\hat{\tilde{Y}}$ as
	%%%%%%%%%%%%%%%%%%%%%%%%%%%%%%%%%%%%%%%%%%%%%%
	\begin{align}
	e^{-2\pi iM\hat{\widetilde{Y}}}|\tilde{y}\rangle= e^{-2\pi iM\tilde{y}}|\tilde{y}\rangle.
	\label{constraint_til_y}
	\end{align}
	%%%%%%%%%%%%%%%%%%%%%%%%%%%%%%%%%%%%%%%%%%%%%%
Making $e^{-i2\pi M\hat{\widetilde{Y}}}$ operate on $e^{ia\hat{\widetilde{P}}}|\tilde{y}\rangle~(a\in \mathbb{R})$ and using eq.(\ref{com_T2_YP}), we can obtain the condition
    %%%%%%%%%%%%%%%%%%%%%%%%%%%%%%%%%%%%%%%%%%%%%%
    \begin{align}
    e^{ia\hat{\widetilde{P}}}|\tilde{y}\rangle = |\tilde{y}-a\rangle.
    \label{ePy_y-1}
    \end{align}
    %%%%%%%%%%%%%%%%%%%%%%%%%%%%%%%%%%%%%%%%%%%%%%
From eqs.(\ref{constraint1}) and (\ref{ePy_y-1}), we can obtain the periodic condition
	%%%%%%%%%%%%%%%%%%%%%%%%%%%%%%%%%%%%%%%%%%%%%%
	\begin{align}
	&|\tilde{y}-1\rangle=|\tilde{y}\rangle ,%~~\tilde{y}=\frac{l}{Q_{12}}~(l\in\mathbb{Z}),
    \label{quantization1}
	\end{align}
	%%%%%%%%%%%%%%%%%%%%%%%%%%%%%%%%%%%%%%%%%%%%%%
with the coordinate quantization condition
	%%%%%%%%%%%%%%%%%%%%%%%%%%%%%%%%%%%%%%%%%%%%%%
	\begin{align}
	\tilde{y}=\frac{j}{M}, ~~~~~(j=0,1,2,\cdots,|M|-1).
    \label{coordinatequantization}
	\end{align}
	%%%%%%%%%%%%%%%%%%%%%%%%%%%%%%%%%%%%%%%%%%%%%%
A schematic figure of the quantization condition is represented in Fig.\ref{fig:one}. 
These results imply that the eigenstates of $\hat{H}$ and their energy eigenvalues are given by 
	%%%%%%%%%%%%%%%%%%%%%%%%%%%%%%%%%%%%%%%%%%%%%%
	\begin{align}
	\hat{H} \left|n,\frac{j}{M}\right\rangle_{T^2} =E_n \left|n,\frac{j}{M}\right\rangle_{T^2},~~~~~~
	E_{n}=\omega \left(n+\frac{1}{2}\right) ,
	\end{align}
	%%%%%%%%%%%%%%%%%%%%%%%%%%%%%%%%%%%%%%%%%%%%%%
where $n=0,1,2,\cdots$ and $j=0,1,2,\cdots,|M|-1$.
Thus, there is $|M|$-fold degeneracy at each energy level in this system, i.e.
	%%%%%%%%%%%%%%%%%%%%%%%%%%%%%%%%%%%%%%%%%%%%%%
	\begin{align}
	{\rm the~number~of~degeneracy}=|M|.
	\end{align}
	%%%%%%%%%%%%%%%%%%%%%%%%%%%%%%%%%%%%%%%%%%%%%%
Note that $E_n$ correspond to eigenvalues of 
two-dimensional Laplace operator with magnetic flux (\ref{Schrodinger}), 
i.e. the mass squares of scalar fields $m^2_n= \omega (n + 1/2)$.
The spinor and two-dimensional vector have 
mass spectra as $m^2_n= \omega n$ and $m^2_n= \omega (n-1/2)$, 
respectively \cite{Cremades:2004wa,Hamada:2012wj}.

	%%%%%%%%%%       Insert the figure here !!        %%%%%%%%%%
	%%%%%%%%%%%%%%%%%%%%%%%%%%%%%%%%%%%

\subsubsection{$Z_{N}$ shifted orbifolding\label{ZNso_T2}}

　From here, we investigate the quantum system on $T^2/Z_{N}$ shifted orbifold, which is defined as the identification
	%%%%%%%%%%%%%%%%%%%%%%%%%%%%%%%%%%%%%%%%%%%%%%
	\begin{align}
	Z_{N}:{\bm y} \sim {\bm y}+\frac{1}{N}{\bm u}_{1}, %~~~~~(N\in \mathbb{N}).
    \label{simpleZN}
	\end{align}
	%%%%%%%%%%%%%%%%%%%%%%%%%%%%%%%%%%%%%%%%%%%%%%
where $N$ is some positive integer.
We should give some comments here that eq.(\ref{simpleZN}) is not the general form in the $Z_N$ shifted orbifolding, but we can always transform the general form into such a special one as eq.(\ref{simpleZN}) without any loss of generality. We show the details in Appendix \ref{soabt}. This identification can be translated into the requirement for physical states on $T^2/Z_{N}$ in the operator formalism as\footnote
{We note that there is always a $Z_{N}$-phase ambiguity to put $\mathrm{e}^{2\pi i \theta/N}~(\theta \in \mathbb{Z})$ in the definition of $\hat{U}_{Z_{N}}$.
Here we assume $\mathrm{e}^{2\pi i \theta/N}=1$ because it does not affect the analysis below.
For the same reason, we will apply this assumption to the cases of other orbifolds.}
 	%%%%%%%%%%%%%%%%%%%%%%%%%%%%%%%%%%%%%%%%%%%%%%
	\begin{align}
	&\hat{U}_{Z_{N}}|\psi\rangle_{T^2/Z_{N}}=|\psi\rangle_{T^2/Z_{N}},~~~~~%\notag \\
	\hat{U}_{Z_{N}}\equiv e^{i\hat{T}_{1}/N},
	\end{align}
	%%%%%%%%%%%%%%%%%%%%%%%%%%%%%%%%%%%%%%%%%%%%%%
where $\hat{U}_{Z_{N}}$ is the operator which shifts the coordinate $\bm{y}$ by ${1\over N}\bm{u}_1$ as in eq.(\ref{simpleZN}).
The $Z_{N}$ shift operator has to be consistent with the torus identification (\ref{constraint}) and leads to the consistency conditions
	%%%%%%%%%%%%%%%%%%%%%%%%%%%%%%%%%%%%%%%%%%%%%%
	\begin{align}
	\hat{U}_{Z_{N}}e^{i\hat{T}_{a}}=e^{i\hat{T}_{a}}\hat{U}_{Z_{N}},~~~~~(a=1,2).
	\end{align}
	%%%%%%%%%%%%%%%%%%%%%%%%%%%%%%%%%%%%%%%%%%%%%%
From these conditions, we can obtain  
	%%%%%%%%%%%%%%%%%%%%%%%%%%%%%%%%%%%%%%%%%%%%%%
	\begin{align}
	\frac{M}{N}\in \mathbb{Z}.
	\end{align}
	%%%%%%%%%%%%%%%%%%%%%%%%%%%%%%%%%%%%%%%%%%%%%%
In other words, we can write
	%%%%%%%%%%%%%%%%%%%%%%%%%%%%%%%%%%%%%%%%%%%%%%
	\begin{align}
	M=tN,
	\label{M=tN}
	\end{align}
	%%%%%%%%%%%%%%%%%%%%%%%%%%%%%%%%%%%%%%%%%%%%%%
where $t$ is some integer.
We can understand briefly what this result means physically. Since the $Z_{N}$ shifted orbifolding reduces the fundamental region $\mathcal{A}$ of $T^2$ to $\mathcal{A}/N$, the magnetic flux quantization requires that $M$ has to be a multiple of $N$. 

Then we define the physical states on $T^2/Z_{N}$ as
	%%%%%%%%%%%%%%%%%%%%%%%%%%%%%%%%%%%%%%%%%%%%%%
	\begin{align}
	\biggl|n,\frac{j}{M}\biggl\rangle_{T^2/Z_{N}}
    &\equiv \frac{1}{\sqrt{N}}\sum^{N-1}_{\ell =0}(\hat{U}_{Z_{N}})^{-\ell}\biggl|n,\frac{j}{M}\biggl\rangle_{T^2}\nonumber\\
	&=\frac{1}{\sqrt{N}}\sum^{N-1}_{\ell =0}\biggl| n,\frac{j+\ell t}{M}\biggl\rangle_{T^2},
	\end{align}
	%%%%%%%%%%%%%%%%%%%%%%%%%%%%%%%%%%%%%%%%%%%%%%
where $j+\ell t$ is identical with an element of the set of $j$ modulo $|M|$.
Here we used the fact that $\hat{U}_{Z_N}$ operate on the states on $T^2$, i.e.
	%%%%%%%%%%%%%%%%%%%%%%%%%%%%%%%%%%%%%%%%%%%%%%
	\begin{align}
	(\hat{U}_{Z_{N}})^{-1}\biggl|n,\frac{j}{M}\biggl\rangle_{T^2}&=\biggl|n,\frac{j}{M}+\frac{1}{N}\biggl\rangle_{T^2}
	=\biggl|n,\frac{j+t}{M}\biggl\rangle_{T^2}.
	\end{align}
	%%%%%%%%%%%%%%%%%%%%%%%%%%%%%%%%%%%%%%%%%%%%%%	

Finally, we mention the number of degeneracy in this system, which is given by 
	%%%%%%%%%%%%%%%%%%%%%%%%%%%%%%%%%%%%%%%%%%%%%%
	\begin{align}
	{\rm the~number~of~degeneracy}=\frac{|M|}{N}=|t|.
	\end{align}
	%%%%%%%%%%%%%%%%%%%%%%%%%%%%%%%%%%%%%%%%%%%%%%
We notice that we cannot obtain any constraint for the number of degeneracy on $T^2/Z_N$.

\subsection{$(T^2\times T^2)/Z_{N}$ shifted orbifold}

In this subsection, using the analysis in subsection \ref{OP_T2ZN}, we
analyze the quantum mechanics on $(T^2
\times T^2)/Z_{N}$ shifted orbifold with a homogeneous magnetic field. 
The extension of the previous analysis is straightforward but the result is nontrivial. 
In this case, the number of degeneracy is restricted to a multiple of $N$. 

\subsubsection{Operator formalism on $T^2\times T^2$}

Let us start from a Hamiltonian of the quantum system on $T^2\times T^2$ with a homogeneous magnetic field, which is given by
	%%%%%%%%%%%%%%%%%%%%%%%%%%%%%%%%%%%%%%%%%%%%%%
	\begin{align}
	H=\sum^{2}_{g=1}\Bigl(-i{\bm \nabla}^{(g)}+q{\bm A}^{(g)}({\bm y}^{(g)})\Bigr)^2,
	\end{align}
	%%%%%%%%%%%%%%%%%%%%%%%%%%%%%%%%%%%%%%%%%%%%%%
where 
	%%%%%%%%%%%%%%%%%%%%%%%%%%%%%%%%%%%%%%%%%%%%%%
	\begin{align}
	{\bm A}^{(g)}({\bm y}^{(g)})=-\frac{1}{2}\Omega^{(g)} {\bm y}^{(g)}+\frac{2\pi}{q}{\bm a}^{(g)},~~~~~(g=1,2). 
    \label{T2T2gaugepotential}
	\end{align}
	%%%%%%%%%%%%%%%%%%%%%%%%%%%%%%%%%%%%%%%%%%%%%%
Here we introduced an index $g$ to be a label of each torus.
The two two-dimensional tori $T^2\times T^{2}$ are defined as
	%%%%%%%%%%%%%%%%%%%%%%%%%%%%%%%%%%%%%%%%%%%%%%
	\begin{align}
	T^2\times T^2 = \mathbb{R}/ \Lambda^{(1)} \times \mathbb{R}/ \Lambda^{(2)},
	\end{align}
	%%%%%%%%%%%%%%%%%%%%%%%%%%%%%%%%%%%%%%%%%%%%%%	
where
	%%%%%%%%%%%%%%%%%%%%%%%%%%%%%%%%%%%%%%%%%%%%%%
	\begin{align}
	\Lambda^{(g)}=\Bigl\{  \hspace{0.2em} \sum^{2}_{a=1}n^{(g)}_{a}{\bm u}^{(g)}_{a} \hspace{0.2em}\Bigl|\hspace{0.2em} n^{(g)}_{a}\in \mathbb{Z} \Bigr\},~~~~~(g=1,2).
	\end{align}
	%%%%%%%%%%%%%%%%%%%%%%%%%%%%%%%%%%%%%%%%%%%%%%
In other words, $T^2\times T^2$ can be defined with the identifications,
	%%%%%%%%%%%%%%%%%%%%%%%%%%%%%%%%%%%%%%%%%%%%%%
	\begin{align}
	{\bm y}^{(g)}\sim {\bm y}^{(g)}+\sum^{2}_{a=1}n_{a}^{(g)}{\bm u}^{(g)}_{a}. %,\hspace{1.3em}( {}^{\forall} n_{a}^{(g)}\in\mathbb{Z}). 
    \label{torusT2T2}
	\end{align}
	%%%%%%%%%%%%%%%%%%%%%%%%%%%%%%%%%%%%%%%%%%%%%%

On each torus, the wave function $\psi({\bm y}^{(1)},{\bm y}^{(2)})$ has to satisfy the pseudo-periodic boundary conditions
 	%%%%%%%%%%%%%%%%%%%%%%%%%%%%%%%%%%%%%%%%%%%%%%
	\begin{align}
	\psi({\bm y}^{(1)}+{\bm u}^{(1)}_{a},{\bm y}^{(2)})&= \mathrm{e}^{iq({\bm y}^{(1)})^{\mathrm{T}}\Omega^{(1)}{\bm u}^{(1)}_{a}/2}\psi({\bm y}^{(1)},{\bm y}^{(2)}), \notag \\
	\psi({\bm y}^{(1)},{\bm y}^{(2)}+{\bm u}_{a}^{(2)})&= \mathrm{e}^{iq({\bm y}^{(2)})^{\mathrm{T}}\Omega^{(2)}{\bm u}^{(2)}_{a}/2}\psi({\bm y}^{(1)},{\bm y}^{(2)}),
    \label{T2T2pseudoperiodic2}
	\end{align}
	%%%%%%%%%%%%%%%%%%%%%%%%%%%%%%%%%%%%%%%%%%%%%%
for the Schr\"odinger equation to be compatible with eq.(\ref{torusT2T2}). The requirement that the wave function is single-valued on each torus leads to the magnetic flux quantization conditions
	%%%%%%%%%%%%%%%%%%%%%%%%%%%%%%%%%%%%%%%%%%%%%%
	\begin{align}
	q{\bm u}^{(g)\mathrm{T}}_{a}B^{(g)}{\bm u}^{(g)}_{b}=2\pi Q_{ab}^{(g)},
	\end{align}
	%%%%%%%%%%%%%%%%%%%%%%%%%%%%%%%%%%%%%%%%%%%%%%
where $B^{(g)}\equiv \frac{1}{2}(\Omega^{(g)}-\Omega^{(g)\mathrm{T}})$ and $Q_{ab}^{(g)}=-Q_{ba}^{(g)}\in\mathbb{Z}$.

Here we introduce momentum operators as ${\bm p}^{(g)}\equiv -i\bm{\nabla}^{(g)}$, which are canonically conjugate to ${\bm y}^{(g)}$. These operators satisfy the canonical commutation relations
	%%%%%%%%%%%%%%%%%%%%%%%%%%%%%%%%%%%%%%%%%%%%%%
	\begin{align}
	[ y^{(g)}_{j},p^{(g')}_{k}] =i\delta_{gg'}\delta_{jk}, %\hspace{1.3em}(j,k=1,2; g,g'=1,2),\\
	~~{\rm the~others}=0,
	\end{align}
	%%%%%%%%%%%%%%%%%%%%%%%%%%%%%%%%%%%%%%%%%%%%%%
where $g,g'=1,2$ and $j,k=1,2$.
In the same way as the previous subsection, the pseudo-periodic boundary conditions (\ref{T2T2pseudoperiodic2}) can be regarded as the constraint conditions for the physical states in the operator formalism. The system is rewritten by
 	%%%%%%%%%%%%%%%%%%%%%%%%%%%%%%%%%%%%%%%%%%%%%%
	\begin{align}
	&\hat{H}|\psi\rangle=E|\psi\rangle, \notag \\
	&\hat{H}=\sum^{2}_{g=1}\left(\hat{{\bm
              p}}^{(g)}-\frac{q}{2}\Omega^{(g)}\hat{{\bm
              y}}^{(g)}+2\pi{\bm a}^{(g)}\right)^2\equiv
        \sum^{2}_{g=1}\left(\hat{{\bm p}}'{}^{(g)}\right)^{2} ,
	\end{align}
	%%%%%%%%%%%%%%%%%%%%%%%%%%%%%%%%%%%%%%%%%%%%%%
with the constraint conditions
 	%%%%%%%%%%%%%%%%%%%%%%%%%%%%%%%%%%%%%%%%%%%%%%
	\begin{align}
	&e^{i\hat{T}_{a}^{(g)}}|\psi\rangle=|\psi\rangle, ~~~~~%\notag \\%~~~~~(a=1,2;g=1,2),\notag \\
	\hat{T}^{(g)}_{a}={\bm u}_{a}^{(g)\mathrm{T}}\left( \hat{{\bm p}}^{(g)}-\frac{q}{2}\Omega^{(g)\mathrm{T}}\hat{{\bm y}}^{(g)} \right), % ~~~(a=1,2;g=1,2).
	\label{T2T2constraint}
    \end{align}
	%%%%%%%%%%%%%%%%%%%%%%%%%%%%%%%%%%%%%%%%%%%%%%
where $a=1,2$ and $g=1,2$.
In the above, we chose the gauge as ${\bm u}_{a}^{(g)\mathrm{T}}\Omega^{(g)}{\bm u}_{a}^{(g)}=0$.
Under the system, we consider the canonical transformation
	%%%%%%%%%%%%%%%%%%%%%%%%%%%%%%%%%%%%%%%%%%%%%%
	\begin{align}
	&\hat{Y}^{(g)} \equiv \frac{\sqrt{2}}{\omega^{(g)}}\hat{p}'{}_{2}^{(g)},~~
	\hat{P}^{(g)}\equiv \sqrt{2}\hat{p}'{}_{1}^{(g)}, \notag \\
	&\hat{\widetilde{Y}}^{(g)}\equiv -\frac{1}{2\pi M^{(g)}}\hat{T}_{2}^{(g)},~~
	\hat{\widetilde{P}}^{(g)}\equiv \hat{T}_{1}^{(g)},
	\end{align}
	%%%%%%%%%%%%%%%%%%%%%%%%%%%%%%%%%%%%%%%%%%%%%%
where $\omega^{(g)}\equiv 2qB^{(g)}_{12}=2qb^{(g)}/|\bm{u}^{(g)}_1\times \bm{u}^{(g)}_2|$ and $M^{(g)}\equiv Q_{12}^{(g)}$.
These new variables satisfy the relations
	%%%%%%%%%%%%%%%%%%%%%%%%%%%%%%%%%%%%%%%%%%%%%%
	\begin{align}	
	[\hat{Y}^{(g)} ,\hat{P}^{(g')} ]=i\delta_{gg'},~~
	[\hat{\widetilde{Y}}{}^{(g)} ,\hat{\widetilde{P}}{}^{(g')} ]=i\delta_{gg'},~~
	{\rm the~others}=0.
	\end{align}
	%%%%%%%%%%%%%%%%%%%%%%%%%%%%%%%%%%%%%%%%%%%%%%
Then, we can rewrite the Hamiltonian as
	%%%%%%%%%%%%%%%%%%%%%%%%%%%%%%%%%%%%%%%%%%%%%%
	\begin{align}	
	&\hat{H}=\sum^{2}_{g=1}\left[
          \frac{1}{2}(\hat{P}^{(g)})^2+\frac{(\omega^{(g)})^2}{2}(\hat{Y}^{(g)})^2
        \right] ,
    \label{T2T2operatorHamiltonian}
    \end{align}
    %%%%%%%%%%%%%%%%%%%%%%%%%%%%%%%%%%%%%%%%%%%%%
with the constraints
    %%%%%%%%%%%%%%%%%%%%%%%%%%%%%%%%%%%%%%%%%%%%%
    \begin{align}
	&e^{i\hat{\widetilde{P}}{}^{(g)}}|\psi\rangle=|\psi\rangle,~~
	e^{-i2\pi M^{(g)}\hat{\widetilde{Y}}{}^{(g)}}|\psi\rangle=|\psi\rangle ~~~~~(g=1,2).
    \label{T2T2operatorconstraint1}
	\end{align}
	%%%%%%%%%%%%%%%%%%%%%%%%%%%%%%%%%%%%%%%%%%%%%%
In a way similar to the previous subsection, $\hat{H}$ is
constructed only from $\hat{P}^{(g)}$ and $\hat{Y}^{(g)}$, so that the
constraint conditions (\ref{T2T2operatorconstraint1}) do not
affect  the energy spectrum of the system. $\hat{H}$ is just the
sum of the two one-dimensional harmonic oscillators, which implies that the energy eigenvalues of $\hat{H}$ are given by
	%%%%%%%%%%%%%%%%%%%%%%%%%%%%%%%%%%%%%%%%%%%%%%
	\begin{align}	
	E_{n^{(1)}n^{(2)}}=\sum^{2}_{g=1}\omega^{(g)}\left(n^{(g)}+\frac{1}{2}\right), ~~~~~(n^{(g)}=0,1,2,\cdots ).
	\end{align}
	%%%%%%%%%%%%%%%%%%%%%%%%%%%%%%%%%%%%%%%%%%%%%%
Furthermore, the constraint conditions (\ref{T2T2operatorconstraint1}) lead to the quantization on the coordinates $\tilde{y}^{(g)}$, 
	%%%%%%%%%%%%%%%%%%%%%%%%%%%%%%%%%%%%%%%%%%%%%%
	\begin{align}	
	\tilde{y}^{(g)}=\frac{j_g}{M^{(g)}}, ~~~~~(j_g=0,1,2,\cdots,|M^{(g)}|-1),
	\end{align}
	%%%%%%%%%%%%%%%%%%%%%%%%%%%%%%%%%%%%%%%%%%%%%%
where $\tilde{y}^{(g)}$ is an eigenvalue of $\hat{\tilde{Y}}^{(g)}$ with the identification $\tilde{y}^{(g)}\sim \tilde{y}^{(g)}+1$.
Thus the eigenstates of this system are described by not only $n^{(1)}$ and $n^{(2)}$ but also $j_1$ and $j_2$ as
 	%%%%%%%%%%%%%%%%%%%%%%%%%%%%%%%%%%%%%%%%%%%%%%
	\begin{align}	
	\biggl| n^{(1)},n^{(2)}; \frac{j_1}{M^{(1)}},\frac{j_2}{M^{(2)}}\biggr\rangle_{T^2\times T^2}. %\hspace{1.3em}(n^{(g)}=0,1,2,\cdots;\hspace{0.2em} l^{(g)}=0,1,2,\cdots,|Q_{12}^{(g)}|-1;\hspace{0.2em} g=1,2).
	\end{align}
	%%%%%%%%%%%%%%%%%%%%%%%%%%%%%%%%%%%%%%%%%%%%%%
Since the eigenvalues of the Hamiltonian are independent of the value $j_g$, there is $|M^{(1)}M^{(2)}|$-fold degeneracy at each energy level, i.e.
	%%%%%%%%%%%%%%%%%%%%%%%%%%%%%%%%%%%%%%%%%%%%%%
	\begin{align}
	{\rm the~number~of~degeneracy }=|M^{(1)}M^{(2)}|.
	\end{align}
	%%%%%%%%%%%%%%%%%%%%%%%%%%%%%%%%%%%%%%%%%%%%%%

\subsubsection{$Z_{N}$ shifted orbifolding}

From here, we investigate the quantum system on $(T^2 \times T^2)/Z_{N}$ shifted orbifold, which is defined by the identification 
	%%%%%%%%%%%%%%%%%%%%%%%%%%%%%%%%%%%%%%%%%%%%%%
	\begin{align}
	Z_{N}:({\bm y}^{(1)},{\bm y}^{(2)})\sim \left({\bm y}^{(1)}+\frac{d}{N}{\bm u}_{1}^{(1)}, {\bm y}^{(2)}+\frac{1}{N}{\bm u}^{(2)}_{1}\right),
    \label{T2T2Znidentification}
	\end{align}
	%%%%%%%%%%%%%%%%%%%%%%%%%%%%%%%%%%%%%%%%%%%%%%
where each of $N$ and $d$ is some integer and $d$ is relatively prime with $N$.
As we mentioned in the previous subsection, eq.(\ref{T2T2Znidentification}) is not the general form in the $Z_N$ shifted orbifolding. 
However, we can always transform the general form into such a special one as eq.(\ref{T2T2Znidentification}) without any loss of generality. 
We discuss it in Appendix \ref{soabt}. 
This identification can be translated into the requirement for physical states in the operator formalism as
	%%%%%%%%%%%%%%%%%%%%%%%%%%%%%%%%%%%%%%%%%%%%%%
	\begin{align}
	\hat{U}_{Z_{N}}|\psi\rangle_{(T^2 \times T^2)/Z_{N}} =|\psi\rangle_{(T^2 \times T^2)/Z_{N}},~~~
    \hat{U}_{Z_{N}}\equiv e^{i(d\hat{T}^{(1)}_{1}+\hat{T}^{(2)}_{1})/N}.
	\end{align}
	%%%%%%%%%%%%%%%%%%%%%%%%%%%%%%%%%%%%%%%%%%%%%%
Since the $Z_{N}$ shift operator has to be compatible with eq.(\ref{T2T2constraint}), we obtain the consistency conditions
	%%%%%%%%%%%%%%%%%%%%%%%%%%%%%%%%%%%%%%%%%%%%%%
	\begin{align}
	\hat{U}_{Z_{N}}\mathrm{e}^{i\hat{T}^{(g)}_{a}}= \mathrm{e}^{i\hat{T}^{(g)}_{a}}\hat{U}_{Z_{N}},~~~~~(a=1,2;g=1,2).
	\end{align}
	%%%%%%%%%%%%%%%%%%%%%%%%%%%%%%%%%%%%%%%%%%%%%%
These conditions lead to 
	%%%%%%%%%%%%%%%%%%%%%%%%%%%%%%%%%%%%%%%%%%%%%%
	\begin{align}
	\frac{dM^{(1)}}{N},~\frac{M^{(2)}}{N}\in\mathbb{Z},
	\end{align}
	%%%%%%%%%%%%%%%%%%%%%%%%%%%%%%%%%%%%%%%%%%%%%%
which imply that both $M^{(1)}$ and $M^{(2)}$ must be multiples of $N$, i.e.
	%%%%%%%%%%%%%%%%%%%%%%%%%%%%%%%%%%%%%%%%%%%%%%
	\begin{align}
	M^{(1)}=t_1N,~~M^{(2)}=t_2N,
	\label{T2T2/ZN_M1M2}
	\end{align}
	%%%%%%%%%%%%%%%%%%%%%%%%%%%%%%%%%%%%%%%%%%%%%%
where each of $t_1$ and $t_2$ is some integer.
 The physical states on $(T^2\times T^2)/Z_{N}$, which are nothing but $Z_{N}$-invariant states, can be constructed as
	%%%%%%%%%%%%%%%%%%%%%%%%%%%%%%%%%%%%%%%%%%%%%%
	\begin{align}
	\biggl|n^{(1)},n^{(2)}; \frac{j_1}{M^{(1)}},\frac{j_2}{M^{(2)}}\biggr\rangle_{(T^2 \times T^2)/Z_{N}} %\notag \\
    &\equiv \frac{1}{\sqrt{N}}\sum_{\ell =0}^{N-1}(\hat{U}_{Z_{N}})^{-\ell}\biggl|n^{(1)},n^{(2)}; \frac{j_1}{M^{(1)}},\frac{j_2}{M^{(2)}}\biggr\rangle_{T^2\times T^2}\nonumber\\
	&= \frac{1}{\sqrt{N}}\sum^{N-1}_{\ell =0}\biggl|n^{(1)},n^{(2)}; \frac{j_1+\ell dt_1}{M^{(1)}},\frac{j_2+\ell t_2}{M^{(2)}}\biggr\rangle_{T^2\times T^2}, \notag \\
	\end{align}
	%%%%%%%%%%%%%%%%%%%%%%%%%%%%%%%%%%%%%%%%%%%%%%
where each of $j_1+\ell dt_1$ and $j_2+\ell t_2$ is identical with an element of the set of $j_g$ modulo $|M^{(g)}|$. 
Moreover, we can obtain the result that the number of degeneracy in this system is a multiple of $N$, i.e.
	%%%%%%%%%%%%%%%%%%%%%%%%%%%%%%%%%%%%%%%%%%%%%%
	\begin{align}
	{\rm the~number~of~degeneracy}=\frac{|M^{(1)}M^{(2)}|}{N}=|t_1t_2|N.
	\label{T2T2/ZN_num_deg}
	\end{align}
	%%%%%%%%%%%%%%%%%%%%%%%%%%%%%%%%%%%%%%%%%%%%%%

We would like to notice that the number of zero-mode fermions for $(T^2\times T^2)/Z_N$ is given by a multiple of $N$, while it can be an arbitrary integer for $T^2/Z_N$.
This result leads to an important conclusion that there is only one possibility to derive the three generations of matter, i.e. $(N;M^{(1)},M^{(2)})=(3;3,3)$ on $(T^2\times T^2)/Z_N$.
Moreover, in a case of $(T^2\times T^2)/Z_N\times T^2$, we obtain only one condition delivering the three generations of matter such as $(N;M^{(1)},M^{(2)},M^{(3)})=(3;3,3,1)$.
We will show that these results coincide with that of the analysis with wave functions in section 3.

\subsection{$(T^2\times T^2\times T^2)/(Z_{N}\times Z_{N'})$ shifted orbifold}

We can also apply the previous results to the system on $(T^2 \times T^2 \times T^2)/(Z_{N}\times Z_{N'})$ shifted orbifold.
%The $Z_{N}$ shift symmetry acts on the first and second tori and the $Z_{N'}$ shift symmetry acts on the socond and third tori. 
As in the previous model, the number of degeneracy is also restricted in this case and can produce three-fold degeneracy. We do not present the analysis by the quantum system on $T^2\times T^2\times T^2$ in detail because it is just a simple extension of the previous one. The energy spectrum of the system on $T^2\times T^2\times T^2$ is labeled by three quantum numbers, $n^{(1)}$, $n^{(2)}$ and $n^{(3)}$, and is given by
	%%%%%%%%%%%%%%%%%%%%%%%%%%%%%%%%%%%%%%%%%%%%%%
	\begin{align}
	E_{n^{(1)}n^{(2)}n^{(3)}}=\sum^{3}_{g=1}\omega^{(g)}\left(n^{(g)}+\frac{1}{2}\right), ~~~~~(n^{(g)}=0,1,2,\cdots),
	\label{E123}
	\end{align}
	%%%%%%%%%%%%%%%%%%%%%%%%%%%%%%%%%%%%%%%%%%%%%%
which is constructed of the three one-dimensional harmonic oscillators.
Then the physical states compatible with eq.(\ref{E123}) are labeled by not only $n^{(g)}$ but also the quantum numbers $j_g$, which come from the constraint conditions in each torus, i.e.
	%%%%%%%%%%%%%%%%%%%%%%%%%%%%%%%%%%%%%%%%%%%%%%
	\begin{align}
	\biggl| n^{(1)},n^{(2)},n^{(3)}; \frac{j_1}{M^{(1)}},\frac{j_2}{M^{(2)}},\frac{j_3}{M^{(3)}}\biggr\rangle_{T^2\times T^2\times T^2},	
	\end{align}
	%%%%%%%%%%%%%%%%%%%%%%%%%%%%%%%%%%%%%%%%%%%%%%
where $j_g=0,1,2,\cdots,|M^{(g)}|-1$.

Next we analyze the quantum system on $(T^2\times T^2\times T^2)/(Z_{N}\times Z_{N'})$. 
We define the $Z_{N}$ and $Z_{N'}$ shifted orbifoldings as the identifications,
	%%%%%%%%%%%%%%%%%%%%%%%%%%%%%%%%%%%%%%%%			
	\begin{align}
	Z_{N}& : ({\bm y}^{(1)},{\bm y}^{(2)},{\bm y}^{(3)})\sim \left({\bm y}^{(1)}+\frac{d}{N}{\bm u}^{(1)}_{1},{\bm y}^{(2)}+\frac{1}{N}{\bm u}^{(2)}_{1},{\bm y}^{(3)}\right), \notag \\%\label{Zny}\\
	Z_{N'}& : ({\bm y}^{(1)},{\bm y}^{(2)},{\bm y}^{(3)})\sim \left({\bm y}^{(1)},{\bm y}^{(2)}+\frac{1}{N'}(s_1{\bm u}^{(2)}_{1}+s_{2}{\bm u}^{(2)}_{2}),{\bm y}^{(3)}+\frac{d'}{N'}{\bm u}^{(3)}_{1}\right),
    \label{Zn'y}
	\end{align}
	%%%%%%%%%%%%%%%%%%%%%%%%%%%%%%%%%%%%%%%%	
where each of $N$, $N'$, $d$ and $d'$ is some positive integer and each of $s_{1}$ and $s_{2}$ is some integer.
Here we require that $d(d')$ is relatively prime with $N(N')$ and when we define $s'$ as the greatest common divisor (g.c.d.) of $s_{1}$ and $s_{2}$, the g.c.d. of $s'$ and $d'$ is relatively prime with $N'$. 
As in Appendix \ref{soabt}, we can always transform the general forms of the $Z_{N}$ and $Z_{N'}$ shifted orbifoldings into such special ones as eq.(\ref{Zn'y}) without any loss of generality.
In the operator formalism, we define the operators generating the identifications (\ref{Zn'y}) as
	%%%%%%%%%%%%%%%%%%%%%%%%%%%%%%%%%%%%%%%%
	\begin{align}
	&\hat{U}_{Z_{N}}\equiv \mathrm{e}^{i(d\hat{T}^{(1)}_{1}+\hat{T}^{(2)}_{1})/N},
    \label{T2T2T2Zn1} \\
	&\hat{U}_{Z_{N'}}\equiv \mathrm{e}^{i\pi s_{1}s_{2}M^{(2)}/N'} \mathrm{e}^{i(s_{1}\hat{T}^{(2)}_{1}+s_{2}\hat{T}^{(2)}_{2}+d'\hat{T}^{(3)}_{1})/N'}.\label{T2T2T2Zn2}
	\end{align}
	%%%%%%%%%%%%%%%%%%%%%%%%%%%%%%%%%%%%%%%%
We note that the phase factor ${e}^{i\pi s_{1}s_{2}M^{(2)}/N'}$ in eq.(\ref{T2T2T2Zn2}) is necessary to be consistent with $(\hat{U}_{Z_{N'}})^{N'}=1$.
The identifications (\ref{Zn'y}) lead to the requirement for physical states on $(T^2\times T^2 \times T^2)/(Z_{N}\times Z_{N'})$ as
	%%%%%%%%%%%%%%%%%%%%%%%%%%%%%%%%%%%%%%%%
	\begin{align}
	&\hat{U}_{Z_{N}}|\psi\rangle_{(T^2\times T^2\times T^2)/(Z_{N}\times Z_{N'})} =|\psi\rangle_{(T^2\times T^2\times T^2)/(Z_{N}\times Z_{N'})},\notag \\
	&\hat{U}_{Z_{N'}}|\psi\rangle_{(T^2\times T^2\times T^2)/(Z_{N}\times Z_{N'})} =|\psi\rangle_{(T^2\times T^2\times T^2)/(Z_{N}\times Z_{N'})}\label{T2T2T2ZnZn2}.
	\end{align}
	%%%%%%%%%%%%%%%%%%%%%%%%%%%%%%%%%%%%%%%%
Since these conditions have to be compatible with the torus conditions
	%%%%%%%%%%%%%%%%%%%%%%%%%%%%%%%%%%%%%%%%
	\begin{align}
	&e^{i\hat{T}^{(g)}_{a}}|\psi\rangle =|\psi\rangle , ~~~~~%\notag \\
	\hat{T}^{(g)}_{a}= {\bm u}^{(g)\mathrm{T}}_{a}\left( \hat{{\bm p}}^{(g)}-\frac{q}{2}\Omega^{(g)\mathrm{T}} \hat{{\bm y}}^{(g)}\right),
	\end{align}
	%%%%%%%%%%%%%%%%%%%%%%%%%%%%%%%%%%%%%%%%	
where $g=1,2,3$ and $a=1,2$, we obtain the consistency conditions
	%%%%%%%%%%%%%%%%%%%%%%%%%%%%%%%%%%%%%%%%
	\begin{align}
	\hat{U}_{Z_{N}}e^{i\hat{T}^{(g)}_{a}}=e^{i\hat{T}^{(g)}_{a}}\hat{U}_{Z_{N}},~~
	\hat{U}_{Z_{N'}}e^{i\hat{T}^{(g)}_{a}}=e^{i\hat{T}^{(g)}_{a}}\hat{U}_{Z_{N'}}.
    \label{T2T2T2constraint2}
	\end{align}
	%%%%%%%%%%%%%%%%%%%%%%%%%%%%%%%%%%%%%%%%
Moreover, the compatibility between the $Z_N$ and $Z_{N'}$ shift operators leads to the extra consistency condition 
	%%%%%%%%%%%%%%%%%%%%%%%%%%%%%%%%%%%%%%%%
	\begin{align}
	&\hat{U}_{Z_{N}}\hat{U}_{Z_{N'}}=\hat{U}_{Z_{N'}}\hat{U}_{Z_{N}}. \label{T2T2T2constraint3}
	\end{align}
	%%%%%%%%%%%%%%%%%%%%%%%%%%%%%%%%%%%%%%%%
From eq.(\ref{T2T2T2constraint2}), we obtain the conditions
	%%%%%%%%%%%%%%%%%%%%%%%%%%%%%%%%%%%%%%%%
	\begin{align}
	&M^{(1)}=t_1N,~~M^{(2)}=t_2 \frac{NN'}{d_2},~~M^{(3)}=t'_3N',
    \label{Q2}
	\end{align}
	%%%%%%%%%%%%%%%%%%%%%%%%%%%%%%%%%%%%%%%%
where each of $t_1$, $t_2$ and $t'_3$ is some integer and $d_2$ is the g.c.d. of $N$ and $N'$. From eq.(\ref{T2T2T2constraint3}), we obtain the additional constraint
	%%%%%%%%%%%%%%%%%%%%%%%%%%%%%%%%%%%%%%%%
	\begin{align}
	\frac{s_{2}M^{(2)}}{NN'}\in\mathbb{Z}.\label{T2T2T2constraint3-1}
	\end{align}
	%%%%%%%%%%%%%%%%%%%%%%%%%%%%%%%%%%%%%%%%
Using eq.(\ref{Q2}), we can rewrite this constraint as
	%%%%%%%%%%%%%%%%%%%%%%%%%%%%%%%%%%%%%%%%
	\begin{align}
	\frac{s_{2}t_2}{d_2}\in\mathbb{Z}.
	\end{align}
	%%%%%%%%%%%%%%%%%%%%%%%%%%%%%%%%%%%%%%%%
When we define the g.c.d. of $s_{2}$ and $d_2$ as $\gamma$,\footnote
{In the case of $s_{2}=0$, we define the g.c.d. of $0$ and $d_2$ as $d_2$.}
 we obtain
	%%%%%%%%%%%%%%%%%%%%%%%%%%%%%%%%%%%%%%%%
	\begin{align}
	s_{2}=\tilde{s}_2 \gamma, ~~d_2=\tilde{d} \gamma,~~t_2=\tilde{t}_2\tilde{d},
	\end{align}
	%%%%%%%%%%%%%%%%%%%%%%%%%%%%%%%%%%%%%%%%
where each of $\tilde{s}_2$ and $\tilde{t}_2$ is some integer and $\tilde{d}$ is some positive integer.
In the case of $(T^2\times T^2\times T^2)/(Z_{N}\times Z_{N'})$, the form of $M^{(2)}$, which is compatible with the consistency conditions, can be rewritten as
	%%%%%%%%%%%%%%%%%%%%%%%%%%%%%%%%%%%%%%%%
	\begin{align}
	M^{(2)}&=\tilde{t}_2\frac{NN'}{\gamma} .
    \label{tNN'/gam}
	\end{align}
	%%%%%%%%%%%%%%%%%%%%%%%%%%%%%%%%%%%%%%%%
 
The physical states on $(T^2\times T^2 \times T^2)/(Z_{N}\times Z_{N'})$, which are nothing but $Z_{N}$- and $Z_{N'}$-invariant states, can be constructed as
	%%%%%%%%%%%%%%%%%%%%%%%%%%%%%%%%%%%%%%%%
	\begin{align}
	\hspace{1.5cm}&\left|n^{(1)},n^{(2)},n^{(3)}; \frac{j_1}{M^{(1)}},\frac{j_2}{M^{(2)}},\frac{j_3}{M^{(3)}}\right\rangle_{(T^2\times T^2 \times T^2)/(Z_{N}\times Z_{N'})}\nonumber\\
	&\equiv \mathcal{N}\sum^{N-1}_{\ell =0} (\hat{U}_{Z_{N}})^{-\ell}\sum^{N'-1}_{\ell' =0} (U_{Z_{N'}})^{-\ell'}\left|n^{(1)},n^{(2)},n^{(3)}; \frac{j_1}{M^{(1)}},\frac{j_2}{M^{(2)}},\frac{j_3}{M^{(3)}}\right\rangle_{T^2 \times T^2 \times T2}\nonumber\\
	&= \mathcal{N}\sum^{N-1}_{\ell =0}\sum^{N'-1}_{\ell' =0}e^{i\pi \ell' s_2(2j_2-(N'-\ell')s_1t'_2)/N'}\nonumber\\
	&\hspace{8mm}\times\left|n^{(1)},n^{(2)},n^{(3)}; \frac{j_1+\ell dt_1}{M^{(1)}},\frac{j_2+\ell t'_2+\ell' s_{1}t''_2}{M^{(2)}},\frac{j_3+\ell'd't'_3}{M^{(3)}}\right\rangle_{T^2 \times T^2 \times T2}, \notag \\
	\end{align}
	%%%%%%%%%%%%%%%%%%%%%%%%%%%%%%%%%%%%%%%%
where $\mathcal{N}$ is the normalization factor, $t'_2\equiv \tilde{t}_2N'/\gamma$ and $t''_2\equiv \tilde{t}_2N/\gamma$.

Then we obtain the result that the number of degeneracy in this system is given by
	%%%%%%%%%%%%%%%%%%%%%%%%%%%%%%%%%%%%%%%%
	\begin{align}
	{\rm the~number~of~degeneracy}=\frac{|M^{(1)}M^{(2)}M^{(3)}|}{NN'}
	=|t_1\tilde{t}_2t'_3| \frac{NN'}{\gamma} ,
	\label{T2T2T2ZNZN_num_deg}
	\end{align}
	%%%%%%%%%%%%%%%%%%%%%%%%%%%%%%%%%%%%%%%%
which is a multiple of $NN'/\gamma$.
Hence, as a conclusion similar to $(T^2\times T^2)/Z_N$, we would like to notice that the number of zero-mode fermions for $(T^2\times T^2\times T^2)/(Z_N\times Z_{N'})$ is given by a multiple of $N$ and $N'$.
This result leads to an important conclusion that there is only one possibility to derive the three generations of matter, i.e.  $(N,N';M^{(1)},M^{(2)},M^{(3)})=(3,3;3,3,3)$ on $(T^2\times T^2 \times T^2)/(Z_N\times Z_{N'})$.
We will show that this result coincides with that of the analysis with wave functions.

In a similar way, we can consider the case of $(T^2\times T^2\times T^2)/(Z_{N}\times Z_{N'}\times Z_{N''})$.
For example, we assume three discrete symmetries as 
\begin{align}
	Z_{N}& : ({\bm y}^{(1)},{\bm y}^{(2)},{\bm y}^{(3)})\sim \Bigl({\bm y}^{(1)}+\frac{1}{N},{\bm y}^{(2)}+\frac{1}{N},{\bm y}^{(3)}\Bigr), \notag \\%\label{Zny}\\
	Z_{N'}& : ({\bm y}^{(1)},{\bm y}^{(2)},{\bm y}^{(3)})\sim \Bigl({\bm y}^{(1)},{\bm y}^{(2)}+\frac{1}{N'},{\bm y}^{(3)}+\frac{1}{N'}\Bigr), \notag \\
    Z_{N''}& : ({\bm y}^{(1)},{\bm y}^{(2)},{\bm y}^{(3)})\sim \Bigl({\bm y}^{(1)}+\frac{1}{N''},{\bm y}^{(2)},{\bm y}^{(3)}+\frac{1}{N''}\Bigr),
    \label{T2T2T2/ZNZNZN_iden}
\end{align}
where each of $N$, $N'$ and $N''$ is some integer. 
Then we obtain the result that the number of degeneracy in this system is given by (see Appendix \ref{digNNN})
\begin{align}
	{\rm the~number~of~degeneracy}&=\frac{|M^{(1)}M^{(2)}M^{(3)}|}{NN'N''} \notag \\
	%&=|t_1t_2t_3| \frac{NN''}{d_1}\cdot \frac{NN'}{d_2}\cdot \frac{N'N''}{d_3}\cdot  \frac{1}{NN'N''} \notag \\
	&=|t_1t_2t_3| \frac{NN'N''}{d_1d_2d_3},
	\label{deg_T6NNN}
\end{align}
where $d_1$ is the g.c.d. of $N$ and $N'$, $d_2$ is the g.c.d. of $N'$ and $N''$, and $d_3$ is the g.c.d. of $N''$ and $N$.
We would also like to notice that there are only two possibilities to derive the three generations of matter, i.e. $(N,N',N'';M^{(1)},M^{(2)},M^{(3)})=(3,9,3;3,9,9),~(3,9,9;9,9,9)$, up to the permutation of parameters for the magnitude of fluxes and the shift symmetries.
We will also show that this result coincides with that of the analysis with wave functions.

%%%%%%%%%%%%%%%%%%%%%%%%%%%%%%%%%%%%%%%%%%%%%%%%%%%%%%%%%%%%%%%%%%%%%%%%%%%%%%%
\section{Wave function on shifted orbifold with magnetic flux}

In section \ref{Op_form}, we considered the fields on shifted orbifolds with magnetic flux by the operator formalism.
Here we re-consider it by analyzing wave functions explicitly because it is necessary to calculate some physical quantities, e.g. Yukawa couplings and higher order couplings \cite{Cremades:2004wa,Abe:2009dr}.
We will see below that the results of the analysis with wave functions coincide with ones by the operator formalism.

\subsection{Review of the $U(1)$ gauge theory on $T^2$\label{RevU1T2}}

\subsubsection{Magnetic flux quantization}

First, we review the $U(1)$ gauge theory on the two-dimensional torus with magnetic flux.\footnote
{This subsection is based on Ref.\cite{Cremades:2004wa}.}
Here, it is convenient to use the complex coordinate $z=y_1+iy_2,~\bar{z}=y_1-iy_2$ instead of the vector notation 
${\bm y}=(y_1,y_2)^{\mathrm{T}}$ in order to write wave functions explicitly.
They satisfy the identification $z\sim z+1\sim z+\tau~(\tau \in \mathbb{C},\mathrm{Im}\tau > 0)$ on $T^2$.\footnote
{For convenience, we choose $(1,\tau)$ as two circumferences of the two-dimensional torus.}%, which to require that the radius of compactification $R$ is equal to $(2\pi)^{-1}$, i.e. $2\pi R=1$.}
~Similarly, we make use of the complex basis for the vector potential as
\begin{align}
A_z={1\over 2}(A_{y_1} -iA_{y_2}),~~A_{\bar{z}}={1\over 2}(A_{y_1}
+iA_{y_2}).
\end{align}
For the non-zero magnetic flux $b$ on $T^2$, we can write that $b=\int_{T^2}F$ by the field strength
\begin{align}
F={ib\over 2\mathrm{Im}\tau}dz\wedge d\bar{z}.
\label{Fb}
\end{align} 
For $F=dA$, the vector potential $A$ can be written as
\begin{align}
A(z,\bar{z})={b\over 2\mathrm{Im}\tau}\mathrm{Im}[(\bar{z}+\bar{a})dz]\equiv A_{z}(z,\bar{z})dz+A_{\bar{z}}(z,\bar{z})d\bar{z},
\label{v_pA}
\end{align}
where $a(\in \mathbb{C})$ is a Wilson line phase.
From eq.(\ref{v_pA}), we obtain 
\begin{align}
&A(z+1,\bar{z}+1)=A(z,\bar{z})+{b\over 2\mathrm{Im}\tau}\mathrm{Im}dz\equiv A(z,\bar{z})+d\chi_1 (z,\bar{z}),~~\notag  \\
&A(z+\tau ,\bar{z}+\bar{\tau})=A(z,\bar{z})+{b\over 2\mathrm{Im}\tau}\mathrm{Im}(\bar{\tau}dz)\equiv A(z,\bar{z})+d\chi_2 (z,\bar{z}),
\end{align}
where\footnote
{For $b\ne 0$, the general functions of $\chi_1(z,\bar{z})$ and $\chi_2(z,\bar{z})$ can be written as 
\begin{align}
\chi_1(z,\bar{z})={b\over 2\mathrm{Im}\tau}\mathrm{Im}(z+a)+{\pi \alpha_1\over q},~~\chi_2(z,\bar{z})={b\over 2\mathrm{Im}\tau}\mathrm{Im}(\bar{\tau}(z+a))+{\pi \alpha_2\over q}, \notag
\end{align}
where $\alpha_1$ and $\alpha_2$ are real numbers.
We can always make $\alpha_i$ absorbed into the Wilson line phase $a$ by the redefinition of fields. (see Appendix \ref{Wlp&alp}.)}
\begin{align}
\chi_1(z,\bar{z})={b\over 2\mathrm{Im}\tau}\mathrm{Im}(z+a),~~\chi_2(z,\bar{z})={b\over 2\mathrm{Im}\tau}\mathrm{Im}(\bar{\tau}(z+a)).
\end{align}
Moreover, let us consider a field $\Phi(z,\bar{z})$ with the $U(1)$ charge $q$ on $T^2$.
We require the Lagrangian density $\mathcal{L}$ to be single-valued as
\begin{align}
\mathcal{L}(A(z,\bar{z}),\Phi(z,\bar{z}))&=\mathcal{L}(A(z+1,\bar{z}+1),\Phi(z+1,\bar{z}+1)) \notag \\
&=\mathcal{L}(A(z+\tau,\bar{z}+\bar{\tau}),\Phi(z+\tau,\bar{z}+\bar{\tau})).
\label{LPhi_T^2}
\end{align}
Then this field $\Phi(z,\bar{z})$ should satisfy the pseudo-periodic boundary conditions
\begin{align}
\Phi(z+1,\bar{z}+1) =e^{iq\chi_1(z,\bar{z})}\Phi(z,\bar{z}),~~\Phi(z+\tau,\bar{z}+\bar{\tau}) =e^{iq\chi_2(z,\bar{z})}\Phi(z,\bar{z}).
\label{Phi_T^2}
\end{align}
From these, the consistency of the contractible loops, e.g. $z\to z+1\to z+1+\tau\to z+\tau\to z$, requires the magnetic flux quantization condition 
\begin{align}
{qb \over 2\pi}\equiv M\in \mathbb{Z}.
\end{align}

\subsubsection{Zero-mode solutions of a fermion}

Here we consider zero-mode solutions of a fermion $\psi (z,\bar{z})$ on $T^2$ with magnetic flux, which satisfy the equation 
\begin{align}
\sum_{a=z,\bar{z}}\Gamma^a(\partial_a -iqA_a)\psi (z,\bar{z})=0,
\label{Deqt2ZN}
\end{align}
where 
\begin{align}
&\partial_z={1\over 2}(\partial_{y_1} -i\partial_{y_2}),~~ \partial_{\bar{z}}={1\over 2}(\partial_{y_1} +i\partial_{y_2}),\notag \\
%&A_z={1\over 2}(A_{y_1} -iA_{y_2}),~~A_{\bar{z}}={1\over 2}(A_{y_1} +iA_{y_2}),% \notag \\ 
&\Gamma^z =\Gamma^1+i\Gamma^2=\sigma^1+i\sigma^2=\left(
\begin{array}{cc}
0&2\\ 0&0
\end{array}
\right),~~\notag \\
&\Gamma^{\bar{z}}=\Gamma^1-i\Gamma^2=\sigma^1-i\sigma^2=\left(
\begin{array}{cc}
0&0\\ 2&0
\end{array}
\right).
\end{align}
Then we can write $\psi (z,\bar{z})$ as a two-component spinor 
\begin{align}
\psi (z,\bar{z})=\left(
\begin{array}{c}
\psi_+ (z,\bar{z})\\ \psi_-(z,\bar{z})
\end{array}
\right),
\end{align}
and eq.(\ref{Deqt2ZN}) can be decomposed as 
\begin{align}
\left(\partial_{\bar{z}} +{\pi M \over 2\mathrm{Im}\tau}(z+a)\right)\psi_+ (z,\bar{z})=0,~~
\left(\partial_z -{\pi M \over 2\mathrm{Im}\tau}(\bar{z}+\bar{a})\right)\psi_- (z,\bar{z})=0.
\label{psi_pm}
\end{align}
The fields $\psi_{\pm}(z,\bar{z})$ should obey the conditions (\ref{Phi_T^2}), i.e. 
\begin{align}
&\psi_\pm (z+1,\bar{z}+1) =e^{iq\chi_1(z,\bar{z})}\psi_\pm
(z,\bar{z}),~~\psi_{\pm}(z+\tau ,\bar{z}+\bar{\tau})
=e^{iq\chi_2(z,\bar{z})}\psi_{\pm} (z,\bar{z}).
\label{psi_trans1tau}
\end{align}
From eqs.(\ref{psi_pm}) and (\ref{psi_trans1tau}), we find that for $M>0~(M<0)$ only $\psi_+~(\psi_-)$ has solutions.
%can obtain the functions of $\psi_{\pm}$.
%If $M > 0$, only $\psi_+$ has zero-modes.
%On the other hand, if $M <0$, only $\psi_-$ has zero-modes.
Their zero-mode wave functions are given by  
\begin{align}
&\psi_+^j (z,\bar{z})=\mathcal{N} e^{i\pi M (z+a){\mathrm{Im}(z+a)\over \mathrm{Im}\tau}}\cdot \vartheta \left[
\begin{array}{c}
{j\over M}\\ 0
\end{array}
\right] (M(z+a),M\tau) ~~~~~{\rm for}~ M>0,
\label{psi_p_func} \\
&\psi_-^j (z,\bar{z})=\mathcal{N} e^{i\pi M (\bar{z}+\bar{a}){\mathrm{Im}(\bar{z}+\bar{a})\over \mathrm{Im}\bar{\tau}}}\cdot \vartheta \left[
\begin{array}{c}
{j\over M}\\ 0
\end{array}
\right] (M(\bar{z}+\bar{a}),M\bar{\tau}) ~~~~~ {\rm for}~ M<0,
\label{psi_m_func}
\end{align}
where $j=0,1,\cdots,|M|-1$ and  $\mathcal{N}$ is the normalization factor.
Here the $\vartheta$-function is defined by 
\begin{align}
&\vartheta \left[
\begin{array}{c}
a\\ b
\end{array}
\right] (\nu,\tau) 
=\sum_{l\in \mathbb{Z}}e^{i\pi (a+l)^2\tau}e^{2\pi i(a+l)(\nu +b)} ,
\end{align}
with the properties
\begin{align}
&\vartheta \left[
\begin{array}{c}
a\\ b
\end{array}
\right] (\nu +n,\tau)
=e^{2\pi i an}\vartheta \left[
\begin{array}{c}
a\\ b
\end{array}
\right] (\nu,\tau), \notag \\
&\vartheta \left[
\begin{array}{c}
a\\ b
\end{array}
\right] (\nu +n\tau,\tau)
=e^{-i\pi n^2\tau -2\pi i n(\nu +b)}\vartheta \left[
\begin{array}{c}
a\\ b
\end{array}
\right] (\nu,\tau),
\end{align}
where $a$ and $b$ are real numbers, $\nu$ and $\tau$ are complex numbers and $\mathrm{Im}\tau >0$.

\subsection{$U(1)$ gauge theory on $T^2/Z_N$ }

Next, we investigate the $U(1)$ gauge theory on $T^2/Z_N$ with magnetic flux, in which the $Z_N$ shifted orbifolding satisfies the identification $z\sim z+e_N^{mn}~(e_N^{mn}\equiv (m+n\tau)/N;m,n\in \mathbb{Z})$.
Here we consider a general $Z_N$ shift $e_N^{mn}$ for convenience of practical computations, although we could take, say, $(m,n)=(1,0)$ without any loss of generality as in eq.(\ref{simpleZN}).
From eq.(\ref{v_pA}), we obtain
\begin{align}
A(z+e_N^{mn},\bar{z}+\bar{e}_N^{mn})=A(z,\bar{z})+{\pi M \over q\mathrm{Im}\tau}\mathrm{Im}(\bar{e}_N^{mn}dz)
\equiv A(z,\bar{z})+d\chi_N(z,\bar{z}).
\label{A_Z_N}
\end{align}
Then we define the physical state $\Phi(z,\bar{z})$ which is consistent with eqs.(\ref{Phi_T^2}) and (\ref{A_Z_N}) as
\begin{align}
&\Phi(z+e_N^{mn},\bar{z}+\bar{e}_N^{mn}) =e^{iq\chi_{N}(z,\bar{z})}\Phi(z,\bar{z}), \label{phys_st} \\
&\chi_{N}(z,\bar{z})={m\over N}\chi_1(z,\bar{z}) +{n\over N}\chi_2(z,\bar{z})+{\pi \alpha_N \over q}, %+{2\pi k\over qN},
%\label{phys_st}
\end{align}
where $\alpha_N$ is some real number and is determined below.
For eq.(\ref{phys_st}) to be consistent with eq.(\ref{Phi_T^2}), we find the relation
\begin{align}
e^{iqN\chi_N(z,\bar{z})}=e^{iq(m\chi_1(z,\bar{z})+n\chi_2(z,\bar{z}))}e^{i\pi mnM}.
\label{N_alpha_M}
\end{align}
It follows that $\alpha_N$ can be determined as $\alpha_N=mnM/N$.
Then the consistency of the contractible loops, e.g. $z\to z+1\to z+1+e_N^{mn}\to z+1+\tau +e_N^{mn}\to z+\tau +e_N^{mn}\to z+e_N^{mn}\to z$, requires the additional conditions such as\footnote
{Actually, we find the consistency conditions  
\begin{align}
e^{i\pi M\left({m\over N}+{n\over N}\right)}=e^{-i\pi M\left({m\over N}+{n\over N}\right)}=e^{i\pi M\left({m\over N}-{n\over N}\right)}=e^{-i\pi M\left({m\over N}-{n\over N}\right)},\notag 
\end{align}
which lead to eq.(\ref{QCT2ZN}).} 
\begin{align}
{mM\over N},~{nM\over N}\in \mathbb{Z}.
\label{QCT2ZN}
\end{align}
Since $N$ has to be relatively prime with the g.c.d. of $m$ and $n$ for $e_N^{mn}$ to represent a $Z_N$ shift symmetry, the conditions (\ref{QCT2ZN}) lead to the magnetic flux quantization condition
\begin{align}
M=tN,
\end{align}
where $t$ is some integer.
We note that this result agrees with eq.(\ref{M=tN}), even though in subsection \ref{ZNso_T2} the magnetic flux quantization condition was derived for a special case of the identification (\ref{simpleZN}), which may correspond to $(m,n)=(1,0)$.

Furthermore, we consider zero-mode fermions $\psi_{\pm}^j (z,\bar{z})$.
It follows from eqs.(\ref{psi_p_func}) and (\ref{psi_m_func}) that $\psi_{\pm}^j (z,\bar{z})$ satisfy the equations 
\begin{align}
\psi_{\pm}^j(z+\ell e_N^{mn},\bar{z}+\ell \bar{e}_N^{mn}) =e^{iq\ell \chi_N(z,\bar{z})}e^{i\pi \ell m(2j-(N-\ell )nt)/N}\psi_{\pm}^{j+\ell nt}(z,\bar{z}), 
\label{psiZ_Ntra}
\end{align}
where $\ell$ is any integer and $\chi_N(z+e_N^{mn},\bar{z}+\bar{e}_N^{mn})=\chi_N(z,\bar{z})$.
Since $\psi_{\pm}^j(z,\bar{z})$ do not, in general, satisfy the physical state condition (\ref{phys_st}) on $T^2/Z_N$, we may need to take appropriate linear combinations of them.
For example, when $(m,n)=(0,1)$,  we obtain 
\begin{align}
\psi_{\pm}^j(z+{\tau\ell\over N},\bar{z}+{\bar{\tau}\ell\over N})
&=e^{iq\ell\chi_N(z,\bar{z})}\psi_{\pm}^{j+\ell t}(z,\bar{z}), 
\end{align}
so the physical states $\Psi_{\pm}^j(z,\bar{z})$ are given by
\begin{align}
\Psi_{\pm}^j(z,\bar{z})
={1\over \sqrt{N}}\sum_{\ell =0}^{N-1}e^{-iq\ell \chi_N(z)}\psi_{\pm}^{j}(z+{\tau\ell\over N},\bar{z}+{\bar{\tau}\ell\over N})
={1\over \sqrt{N}}\sum_{\ell =0}^{N-1}\psi_{\pm}^{j+\ell t}(z,\bar{z}),
\end{align}
where $j=0,1,\cdots,|t|-1$.
We can check that these $\Psi_{\pm}^j(z,\bar{z})$ indeed satisfy eq.(\ref{phys_st}).
When $(m,n)=(1,0)$, we obtain 
\begin{align}
\psi_{\pm}^j(z+{\ell \over N},\bar{z}+{\ell \over N})
&=e^{iq\ell \chi_N(z,\bar{z})}e^{2\pi i j\ell/N}\psi_{\pm}^{j}(z,\bar{z}),
\end{align}
so the physical states $\Psi_{\pm}^j(z,\bar{z})$ are given by 
\begin{align}
\Psi_{\pm}^j(z,\bar{z})
&={1\over N}\sum_{\ell =0}^{N-1}e^{-iq\ell \chi_N(z,\bar{z})}\psi_{\pm}^j(z+{\ell \over N},\bar{z}+{\ell \over N}) \notag \\
&={1\over N}\sum_{\ell =0}^{N-1}e^{2\pi ij\ell/N}\psi_{\pm}^j(z,\bar{z}) 
=\left\{
\begin{array}{cc}
\psi_{\pm}^j(z,\bar{z})&(j\equiv 0~\mathrm{mod}~N) \\ 0&(j\not\equiv 0~\mathrm{mod}~N)
\end{array}
\right..
\end{align}
We can check that these $\Psi_{\pm}^j(z,\bar{z})$ satisfy eq.(\ref{phys_st}).

For a general $Z_N$ shift $e_N^{mn}$, we can obtain the physical states $\Psi_{\pm}^j(z,\bar{z})$, which satisfy eq.(\ref{phys_st}), as
\begin{align}
\Psi_{\pm}^j(z,\bar{z})
&=\mathcal{N}'\sum_{\ell =0}^{N-1}e^{-iq\ell \chi_N(z,\bar{z})}\psi_{\pm}^{j}(z+\ell e_N^{mn},\bar{z}+\ell \bar{e}_N^{mn})\notag \\
&=\mathcal{N}'\sum_{\ell =0}^{N-1}e^{i\pi \ell m(2j-(N-\ell)nt)/N}\psi_{\pm}^{j+\ell nt}(z,\bar{z}),
\label{T2/ZN_ZN_inv_genePsi}
\end{align}
where $\mathcal{N}'$ is the normalization factor.
We notice that we cannot obtain any constraint on the number of zero-mode fermions, i.e. the generation of matter in the standard model, because the number of degeneracy $t$ is a free parameter.

Here, we give a brief comment on couplings on 
$T^2/Z_N$.
On the two-dimensional torus, a generic $L$-point coupling $C^{j_{(1)}\cdots j_{(L)}}$ 
in a 4D low-energy effective field theory
is given by 
the overlap integral of zero-mode functions \cite{Cremades:2004wa,Abe:2009dr},
\begin{align}
C^{j_{(1)}\cdots j_{(L)}}=c^{j_{(1)}\cdots j_{(L)}} 
\int_{T^2} d^2z ~\psi^{j_{(1)}} \cdots \psi^{j_{(L)}},
\end{align}
where $c^{j_{(1)}\cdots j_{(L)}}$ denotes the coupling 
in a higher dimensional field theory.
Similarly, on the orbifold $T^2/Z_N$, a generic 
$L$-point coupling $C^{j_{(1)}\cdots j_{(L)}}_{\rm orbifold}$ 
is given as  
\begin{align}
C^{j_{(1)}\cdots j_{(L)}}_{\rm orbifold}=c^{j_{(1)}\cdots j_{(L)}} 
\int_{T^2/Z_N} d^2z ~\Psi^{j_{(1)}} \cdots \Psi^{j_{(L)}}.
\end{align}
Since the zero-mode wave functions $\Psi^j$ are 
written by linear combinations of $\psi^j$, as shown in 
eq.(\ref{T2/ZN_ZN_inv_genePsi}), 
the coupling $C^{j_{(1)}\cdots j_{(L)}}_{\rm orbifold}$ is 
written by a proper linear combination of $C^{j_{(1)}\cdots j_{(L)}}$.
Its extension to other orbifolds such as $(T^2\times T^2)/Z_N$, 
$(T^2 \times T^2 \times T^2)/(Z_N \times Z_{N'})$ and 
$(T^2 \times T^2 \times T^2)/(Z_N \times Z_{N'} \times Z_{N''})$
is also straightforward.

Although we have constructed the $Z_N$-invariant wave functions $\Psi_{\pm}^j(z,\bar{z})$ as in eq.(\ref{T2/ZN_ZN_inv_genePsi}), it is also worthwhile to consider wave functions $\Phi_{\kappa}(z,\bar{z})$ with a $Z_N$ charge $\kappa$, which is defined as
\begin{align}
&\Phi_{\kappa}(z+e_N^{mn},\bar{z}+\bar{e}_N^{mn}) =\omega^{\kappa } e^{iq\chi_{N}(z,\bar{z})}\Phi_{\kappa}(z,\bar{z}), 
\label{PkP_T2}%\notag \\
%&\chi_{N}(z,\bar{z})={m\over N}\chi_1(z,\bar{z}) +{n\over N}\chi_2(z,\bar{z})+\pi \alpha_N ,
\end{align}
where $\kappa$ is some integer and $\omega \equiv e^{2\pi i/N}$.
Then, in a way similar to eq.(\ref{T2/ZN_ZN_inv_genePsi}), the wave functions $\Psi_{\pm,\kappa}^{j}(z,\bar{z})$ satisfying eq.(\ref{PkP_T2}) can be constructed from $\psi_{\pm}^j(z,\bar{z})$ and are given by
\begin{align}
\Psi_{\pm,\kappa}^j(z,\bar{z})
&=\mathcal{N}'\sum_{\ell =0}^{N-1}\omega^{-\ell \kappa }e^{-iq\ell \chi_N(z,\bar{z})}\psi_{\pm}^{j}(z+\ell e_N^{mn},\bar{z}+\ell \bar{e}_N^{mn})\notag \\
&=\mathcal{N}'\sum_{\ell =0}^{N-1}e^{i\pi \ell m(2j-(N-\ell )nt)/N}e^{-2\pi i\ell \kappa /N}\psi_{\pm}^{j+\ell nt}(z,\bar{z}).
\end{align}

\subsection{$U(1)$ gauge theory on $(T^2\times T^2)/Z_N$}

In a similar way, we consider the $U(1)$ gauge theory on $(T^2\times T^2)/Z_N$ with magnetic flux. 
Let us define $z^{(1)}$ and $z^{(2)}$ as the complex coordinates for each torus with the identifications $z^{(g)}\sim z^{(g)}+1 \sim z^{(g)}+\tau^{(g)}$.
When there are non-zero magnetic fluxes $b^{(1)}$ and $b^{(2)}$ on $T^2\times T^2$, we can write that $b^{(g)}=\int_{T^{2(g)}}F^{(g)}$ by the field strengths 
\begin{align}
F^{(g)}={ib^{(g)}\over 2\mathrm{Im}\tau^{(g)}}dz^{(g)}\wedge d\bar{z}^{(g)}, ~~~~~(g=1,2).
\end{align}  
For $F^{(g)}=dA^{(g)}$, the vector potentials $A^{(g)}$ can be written as\footnote{From here we omit $\bar{z}^{(g)}$ such as $A^{(g)}(z^{(g)})$, $\Phi(z^{(g)})$ and $\psi(z^{(g)})$.}
\begin{align}
A^{(g)}(z^{(g)})={b^{(g)}\over 2\mathrm{Im}\tau^{(g)}}\mathrm{Im}[(\bar{z}^{(g)}+\bar{a}^{(g)})dz^{(g)}]\equiv A_{z^{(g)}}^{(g)}(z^{(g)})dz^{(g)}+A_{\bar{z}^{(g)}}^{(g)}(z^{(g)})d\bar{z}^{(g)},
\label{v_pA2}
\end{align}
where $a^{(g)}(\in \mathbb{C})$ are Wilson line phases.
From eq.(\ref{v_pA2}), we obtain 
\begin{align}
&A^{(g)}(z^{(g)}+1)=A^{(g)}(z^{(g)})+{b^{(g)}\over 2\mathrm{Im}\tau^{(g)}}\mathrm{Im}dz^{(g)}\equiv A^{(g)}(z^{(g)})+d\chi_1^{(g)} (z^{(g)}),~~\notag  \\
&A^{(g)}(z^{(g)}+\tau^{(g)} )=A^{(g)}(z^{(g)})+{b^{(g)}\over 2\mathrm{Im}\tau^{(g)}}\mathrm{Im}(\bar{\tau}^{(g)}dz^{(g)})\equiv A^{(g)}(z^{(g)})+d\chi_2^{(g)} (z^{(g)}),
\end{align}
where
\begin{align}
\chi_1^{(g)}(z^{(g)})={b^{(g)}\over 2\mathrm{Im}\tau^{(g)}}\mathrm{Im}(z^{(g)}+a^{(g)}),~~\chi_2^{(g)}(z^{(g)})={b^{(g)}\over 2\mathrm{Im}\tau^{(g)}}\mathrm{Im}(\bar{\tau}^{(g)}(z^{(g)}+a^{(g)})).
\end{align}
Let us next consider a field $\Phi(z^{(1)},z^{(2)})$ with the $U(1)$ charge $q$ on $T^2\times T^2$.
For the Lagrangian density to be single-valued on $T^2\times T^2$, we require that the field $\Phi(z^{(1)},z^{(2)})$ satisfies the pseudo-periodic boundary conditions
\begin{align}
&\Phi(z^{(1)}+1, z^{(2)}) =e^{iq\chi_1^{(1)}(z^{(1)})}\Phi(z^{(1)} ,z^{(2)}),~~
\Phi(z^{(1)}+\tau^{(1)}, z^{(2)}) =e^{iq\chi_2^{(1)}(z^{(1)})}\Phi(z^{(1)},z^{(2)}), \notag \\
&\Phi(z^{(1)}, z^{(2)}+1) =e^{iq\chi_1^{(2)}(z^{(2)})}\Phi(z^{(1)},z^{(2)}),~~ 
\Phi(z^{(1)},z^{(2)}+\tau^{(2)}) =e^{iq\chi_2^{(2)}(z^{(2)})}\Phi(z^{(1)},z^{(2)}). \notag \\
\label{T^22_Phi}
\end{align}
The compatibility of the conditions (\ref{T^22_Phi}) with any contractible loops requires the magnetic flux quantization conditions 
\begin{align}
{qb^{(1)}\over 2\pi}\equiv M^{(1)},~{qb^{(2)}\over 2\pi}\equiv M^{(2)}\in \mathbb{Z}.
\end{align}

Moreover, we consider $(T^2\times T^2)/Z_N$ to impose a $Z_N$ shift symmetry on $T^2\times T^2$ with the additional identification $(z^{(1)},z^{(2)})\sim (z^{(1)}+e_N^{m_1n_1},z^{(2)}+e_N^{m_2n_2})$.
For the Lagrangian density to be single-valued on $(T^2\times T^2)/Z_N$, we define $\Phi(z^{(1)},z^{(2)})$ following eq.(\ref{T^22_Phi}) as 
\begin{align}
&\Phi(z^{(1)}+e_N^{m_1n_1},z^{(2)}+e_N^{m_2n_2})=e^{iq(\chi_N^{(1)}(z^{(1)})+\chi_N^{(2)}(z^{(2)}))}\Phi(z^{(1)},z^{(2)}),
\label{phys_st2} \\
&\chi_{N}^{(g)}(z^{(g)})={m_g\over N}\chi_1^{(g)}(z^{(g)}) +{n_g\over N}\chi_2^{(g)}(z^{(g)})+{\pi \alpha_{N}^{(g)} \over q} ,%+{2\pi k_g\over qN},
\end{align}
where $\alpha_{N}^{(g)}$ are some real numbers.
For eq.(\ref{phys_st2}) to be consistent with eq.(\ref{T^22_Phi}), we find the relation
\begin{align}
e^{iqN(\chi_N^{(1)}(z^{(1)})+\chi_N^{(2)}(z^{(2)}))}=\prod_{g=1,2}e^{iq(m_g\chi_1^{(g)}(z^{(g)})+n_g\chi_2^{(g)}(z^{(g)}))}e^{i\pi m_gn_gM^{(g)}},
\end{align}
which determines the values of $\alpha_N^{(g)}$ to be $\alpha_{N}^{(g)}=m_gn_gM^{(g)}/N$.
Then the consistency of the contractible loops requires the additional magnetic flux quantization conditions 
\begin{align}
{m_1M^{(1)}\over N},~{n_1M^{(1)}\over N},~{m_2M^{(2)}\over N},~{n_2M^{(2)}\over N}\in \mathbb{Z}.
\end{align}
In the same way as $T^2/Z_N$, each of $M^{(g)}$ turns out to be a multiple of $N$, i.e.
\begin{align}
M^{(1)}=t_1N,~~M^{(2)}=t_2N,
\end{align}
where each of $t_1$ and $t_2$ is some integer.
We note that this result also agrees with eq.(\ref{T2T2/ZN_M1M2}) and the number of degeneracy is given by eq.(\ref{T2T2/ZN_num_deg}).

Next, we consider zero-mode solutions of a fermion $\psi (z^{(1)},z^{(2)})$ on $(T^2\times T^2)/Z_N$, which satisfies the equation 
\begin{align}
\sum_{g=1}^{2}\sum_{a=z^{(g)},\bar{z}^{(g)}}\Gamma^a(\partial_a -iqA_a^{(g)})\psi (z^{(1)},z^{(2)})=0,
\end{align}
where 
\begin{align}
&\Gamma^{z^{(1)}}=\left(
\begin{array}{cc}
 0&2 \\0&0
\end{array}
\right)\otimes \sigma^0,~
\Gamma^{\bar{z}^{(1)}}=\left(
\begin{array}{cc}
 0&0\\ 2&0
\end{array}
\right)\otimes \sigma^0,~\notag \\
&\Gamma^{z^{(2)}}=\sigma^3\otimes \left(
\begin{array}{cc}
  0&2 \\0&0
\end{array}
\right),~
\Gamma^{\bar{z}^{(2)}}=\sigma^3\otimes \left(
\begin{array}{cc}
 0&0\\ 2&0
\end{array}
\right).
\end{align}
Then we can write $\psi(z^{(1)},z^{(2)})$ as a four-component spinor
\begin{align}
\psi (z^{(1)},z^{(2)})=\Biggl(
\begin{array}{c}
 \psi_{+}^{j_1}(z^{(1)})\\ \psi_{-}^{j_1}(z^{(1)})
\end{array}
\Biggr) \otimes
\Biggl(
\begin{array}{c}
 \psi_{+}^{j_2}(z^{(2)})\\ \psi_{-}^{j_2}(z^{(2)})
\end{array}
\Biggr) 
\equiv \Biggl(~\psi_{\mathcal{P}}^J(z^{(1)},z^{(2)})~\Biggr),
\end{align}
where $\mathcal{P}\equiv (\pm,\pm), (\pm,\mp)$ and $J\equiv (j_1,j_2)$.
In the same reason as $T^2$, depending on $M^{(1)}\lessgtr 0$ and $M^{(2)}\lessgtr 0$, only one of $\psi_{\mathcal{P}}$ is well-defined, while the others cannot be normalizable.

Furthermore, the zero-mode fermions $\psi_{\mathcal{P}}^J(z^{(1)},z^{(2)})$ are constructed of $\psi_{\pm}^{j_g}(z^{(g)})$ on each torus, and from eqs.(\ref{psi_p_func}) and (\ref{psi_m_func}), $\psi_{\pm}^{j_g} (z^{(g)})$ satisfy the equations 
\begin{align}
\psi_{\pm}^{j_g}(z^{(g)}+\ell e_N^{m_gn_g}) =e^{iq\ell \chi_N^{(g)}(z^{(g)})}e^{i\pi \ell m_g(2j_g-(N-\ell )n_gt_g)/N}\psi_{\pm}^{j_g+\ell n_gt_g}(z^{(g)}), 
\label{psiZ_Ntra2}
\end{align}
where $\ell$ is any integer and $\chi_N^{(g)}(z^{(g)}+e_N^{m_gn_g})=\chi_N^{(g)}(z^{(g)})$.
Then, since $\psi_{\mathcal{P}}^J(z^{(1)},z^{(2)})$ do not, in general, satisfy the physical state condition (\ref{phys_st2}) on $(T^2\times T^2)/Z_N$, we may need to take appropriate linear combinations of them in order to obtain the physical states $\Psi_{\mathcal{P}}^J(z^{(1)},z^{(2)})$.
For example, when $(m_1,n_1,m_2,n_2)=(0,1,0,1)$, the physical states $\Psi_{\mathcal{P}}^J(z^{(1)},z^{(2)})$ are given by
\begin{align}
\Psi_{\mathcal{P}}^{J}(z^{(1)},z^{(2)})
= {1\over \sqrt{N}}\sum_{\ell =0}^{N-1}\psi_{\mathcal{P}}^{J+\ell T}(z^{(1)},z^{(2)}),
\end{align}
where $J+\ell T\equiv (j_1+\ell t_1,j_2+\ell t_2)$ and the number of degeneracy of $\Psi_{\mathcal{P}}^J(z^{(1)},z^{(2)})$ is $|t_1t_2|N$.
When $(m_1,n_1,m_2,n_2)=(1,0,1,0)$, the physical states $\Psi_{\mathcal{P}}^J(z^{(1)},z^{(2)})$ are given by
\begin{align}
\Psi_{\mathcal{P}}^J(z^{(1)},z^{(2)})
&= {1\over N}\sum_{\ell =0}^{N-1}e^{2\pi i\ell (j_1+j_2)/N}\psi_{\mathcal{P}}^J(z^{(1)},z^{(2)}) \notag \\
&=\left\{
\begin{array}{cc}
\psi_{\mathcal{P}}^J(z^{(1)},z^{(2)})&((j_1+j_2)\equiv 0~\mathrm{mod}~N) \\ 0&((j_1+j_2)\not\equiv 0~\mathrm{mod}~N)
\end{array}
\right.,
\label{eq-phys-state}
\end{align}
where the number of degeneracy of $\Psi_{\mathcal{P}}^J(z^{(1)},z^{(2)})$ is $|t_1t_2|N$.

In the same way, for a general $Z_N$ shift $e_N^{m_gn_g}$ the physical states $\Psi_{\mathcal{P}}^J(z^{(1)},z^{(2)})$, which satisfy eq.(\ref{phys_st2}), are given by
\begin{align}
\Psi_{\mathcal{P}}^{J}(z^{(1)},z^{(2)}) 
=\mathcal{N}'\sum_{\ell =0}^{N-1}\prod_{g=1,2}e^{i\pi \ell m_g(2j_g-(N-\ell)n_gt_g)/N}\psi_{\mathcal{P}}^{J+\ell T_{n_1n_2}}(z^{(1)},z^{(2)}),
\label{T2T2/ZN_ZN_inv_genePsi}
\end{align}
where $\mathcal{N}'$ is the normalization factor and $J+\ell T_{n_1n_2}\equiv (j_1+\ell n_1t_1,j_2+\ell n_2t_2)$.

We would like to notice that the number of zero-mode fermions for $(T^2\times T^2)/Z_N$ is given by a multiple of $N$, while it can be an arbitrary integer for $T^2/Z_N$.
This result coincides with that of the operator formalism and leads to an important conclusion that there is only one possibility to derive the three generations of matter, i.e. $(N;M^{(1)},M^{(2)})=(3;3,3)$ on $(T^2\times T^2)/Z_N$.
Moreover, in a case of $(T^2\times T^2)/Z_N\times T^2$, we obtain only one condition delivering the three generations of matter such as $(N;M^{(1)},M^{(2)},M^{(3)})=(3;3,3,1)$.

Furthermore, it is also worthwhile to consider wave functions $\Phi_{\kappa}(z^{(1)},z^{(2)})$ with a $Z_N$ charge $\kappa$, which is defined by 
\begin{align}
\Phi_{\kappa}(z^{(1)}+e_N^{m_1n_1},z^{(2)}+e_N^{m_2n_2})=\omega^{\kappa }e^{iq(\chi_N^{(1)}(z^{(1)})+\chi_N^{(2)}(z^{(2)}))}\Phi_{\kappa}(z^{(1)},z^{(2)}),
\label{PkP_T2T2}
\end{align}
where $\kappa $ is some integer and $\omega \equiv e^{2\pi i/N}$.
Then the wave functions $\Psi_{\mathcal{P},\kappa}^{J}(z^{(1)},z^{(2)})$ satisfying eq.(\ref{PkP_T2T2}) can be constructed from $\psi_{\pm}^{j_g}(z^{(g)})$ and are given by
\begin{align}
&\Psi_{\mathcal{P},\kappa}^{J}(z^{(1)},z^{(2)}) \notag \\
&=\mathcal{N}'\sum_{\ell =0}^{N-1}\prod_{g=1,2}e^{i\pi \ell m_g(2j_g-(N-\ell)n_gt_g)/N}e^{-2\pi i\ell \kappa /N}\psi_{\mathcal{P}}^{J+\ell T_{n_1n_2}}(z^{(1)},z^{(2)}).
\end{align}

\subsection{$U(1)$ gauge theory on $(T^2\times T^2\times T^2)/(Z_N\times Z_{N'})$}

In a way similar to the case of $(T^2\times T^2)/Z_N$, we consider the $U(1)$ gauge theory on $(T^2\times T^2\times T^2)/(Z_N\times Z_{N'})$ with magnetic flux.
Here we leave the full analysis out and discuss some points.
We impose the $Z_N$ shift symmetry on the first and second tori, which relates the first torus with the second one as $(z^{(1)},z^{(2)},z^{(3)})\sim (z^{(1)}+e_N^{m_1n_1},z^{(2)}+e_N^{m_2n_2},z^{(3)})$.
On the other hand, we impose the $Z_{N'}$ shift symmetry on the second and third tori, which relates the second torus with the third one as $(z^{(1)},z^{(2)},z^{(3)})\sim (z^{(1)},z^{(2)}+e_{N'}^{m'_2n'_2},z^{(3)}+e_{N'}^{m'_3n'_3})$.
In this case, for the Lagrangian density to be single-valued, the pseudo-periodic boundary conditions of the physical states $\Phi(z^{(1)},z^{(2)},z^{(3)})$ with the $U(1)$ charge $q$ are given by
\begin{align}
&\Phi(z^{(1)}+e_N^{m_1n_1},z^{(2)}+e_N^{m_2n_2},z^{(3)})=e^{iq(\chi_N^{(1)}(z^{(1)})+\chi_N^{(2)}(z^{(2)}))}\Phi(z^{(1)},z^{(2)},z^{(3)}), \label{phys_st3_1} \\
&\chi_{N}^{(g)}(z^{(g)})={m_g\over N}\chi_1^{(g)}(z^{(g)}) +{n_g\over N}\chi_2^{(g)}(z^{(g)})+{\pi \alpha_{N}^{(g)} \over q}, % +{2\pi k_g\over qN}, 
\\
&\Phi(z^{(1)},z^{(2)}+e_{N'}^{m'_2n'_2},z^{(3)}+e_{N'}^{m'_3n'_3})=e^{iq(\chi_{N'}^{(2)}(z^{(2)})+\chi_{N'}^{(3)}(z^{(3)}))}\Phi(z^{(1)},z^{(2)},z^{(3)}), \label{phys_st3_2} \\
&{\chi}_{N'}^{(g')}(z^{(g')})={m'_{g'}\over N'}{\chi'}_1^{(g')}(z^{(g')}) +{n'_{g'}\over N'}{\chi'}_2^{(g')}(z^{(g')})+{\pi {\alpha'}_{N'}^{(g')} \over q}, %+{2\pi k'_{g'}\over qN'},
\end{align}
with the relations
\begin{align}
&e^{iqN(\chi_N^{(1)}(z^{(1)})+\chi_N^{(2)}(z^{(2)}))}=\prod_{g=1,2}e^{iq(m_g\chi_1^{(g)}(z^{(g)})+n_g\chi_2^{(g)}(z^{(g)}))}e^{i\pi m_gn_gM^{(g)}},  \\
&e^{iqN'({\chi}_{N'}^{(2)}(z^{(2)})+{\chi}_{N'}^{(3)}(z^{(3)}))}=\prod_{g'=2,3}e^{iq(m'_{g'}{\chi'}_1^{(g')}(z^{(g')})+n'_{g'}{\chi'}_2^{(g')}(z^{(g')}))}e^{i\pi m'_{g'}n'_{g'}M^{(g')}},
\end{align}
where $g=1,2$, $g'=2,3$, $\alpha_{N}^{(g)}=m_gn_gM^{(g)}/N$ and ${\alpha'}_{N'}^{(g')}=m'_{g'}n'_{g'}M^{(g')}/N'$.
Then the consistency of the contractible loops requires the magnetic flux quantization conditions 
\begin{align}
{m_gM^{(g)}\over N},~{n_gM^{(g)}\over N},~{m'_{g'}M^{(g')}\over N'},~{n'_{g'}M^{(g')}\over N'}\in \mathbb{Z}.
\end{align}
From these conditions, each of $M^{(1)}$, $M^{(2)}$ and $M^{(3)}$ turns out to be a multiple of $N$ and/or $N'$, i.e. 
\begin{align}
M^{(1)}=t_1N,~~M^{(2)}=t_2N=t'_2N',~~M^{(3)}=t'_3N',
\end{align}
where each of $t_g$ and $t'_{g'}$ is some integer.
Defining $d$ as the g.c.d. of $N$ and $N'$, we obtain $N\equiv \tilde{n}d$ and $N'\equiv \tilde{n}'d$, where each of $\tilde{n}$ and $\tilde{n}'$ is some positive integer and $\tilde{n}$ is relatively prime with $\tilde{n}'$. 
Since $t_2N=t'_2N'$, we obtain the relation $t_2\tilde{n}=t'_2\tilde{n}'$.
When $|\tilde{t}|$ is defined as the g.c.d. of $t_2$ and $t'_2$, we can rewrite $t_1$ and $t_2$ as $t_2= \tilde{n}'\tilde{t}$ and $t'_2= \tilde{n}\tilde{t}$, respectively.
Namely, we obtain 
\begin{align}
M^{(2)}=\tilde{t}{NN'\over d}.%=\tilde{t}_2{NN'\over \gamma},
\end{align}
If $\tilde{d}$ is defined as the g.c.d. of $\tilde{t}$ and $d$, we obtain the same equation as eq.(\ref{tNN'/gam}),
\begin{align}
M^{(2)}=\tilde{t}_2{NN'\over \gamma},
\end{align}
where each of $\tilde{t}_2$ and $\gamma$ is some positive integer, $\tilde{t}=\tilde{t}_2\tilde{d}$ and $d=\gamma \tilde{d}$.
We note that this result also agrees with eq.(\ref{tNN'/gam}) and the number of degeneracy is given by eq.(\ref{T2T2T2ZNZN_num_deg}).

In a way similar to the case of $(T^2\times T^2)/Z_N$, we also consider zero-mode solutions of a fermion $\psi (z^{(1)},z^{(2)},z^{(3)})$ on $(T^2\times T^2\times T^2)/(Z_N\times Z_{N'})$, which satisfies the equation 
\begin{align}
\sum_{g=1}^{3}\sum_{a=z^{(g)},\bar{z}^{(g)}}\Gamma^a(\partial_a -iqA_a^{(g)})\psi (z^{(1)},z^{(2)},z^{(3)})=0,
\end{align}
where 
\begin{align}
&\Gamma^{z^{(1)}}=\left(
\begin{array}{cc}
 0&2 \\0&0
\end{array}
\right)\otimes \sigma^0\otimes \sigma^0,~
\Gamma^{\bar{z}^{(1)}}=\left(
\begin{array}{cc}
 0&0\\ 2&0
\end{array}
\right)\otimes \sigma^0\otimes \sigma^0,~\notag \\
&\Gamma^{z^{(2)}}=\sigma^3\otimes \left(
\begin{array}{cc}
  0&2 \\0&0
\end{array}
\right)\otimes \sigma^0,~
\Gamma^{\bar{z}^{(2)}}=\sigma^3\otimes \left(
\begin{array}{cc}
 0&0\\ 2&0
\end{array}
\right)\otimes \sigma^0, \notag \\
&\Gamma^{z^{(3)}}=\sigma^3\otimes \sigma^3\otimes \left(
\begin{array}{cc}
  0&2 \\0&0
\end{array}
\right),~
\Gamma^{\bar{z}^{(3)}}=\sigma^3\otimes \sigma^3\otimes \left(
\begin{array}{cc}
 0&0\\ 2&0
\end{array}
\right).
\end{align}
Then we can write $\psi (z_1,z_2,z_3)$ as an eight-component spinor
\begin{align}
\psi (z^{(1)},z^{(2)},z^{(3)})
&=\Biggl(
\begin{array}{c}
 \psi_{+}^{j_1}(z^{(1)})\\ \psi_{-}^{j_1}(z^{(1)})
\end{array}
\Biggr) \otimes
\Biggl(
\begin{array}{c}
 \psi_{+}^{j_2}(z^{(2)})\\ \psi_{-}^{j_2}(z^{(2)})
\end{array}
\Biggr) \otimes
\Biggl(
\begin{array}{c}
 \psi_{+}^{j_3}(z^{(3)})\\ \psi_{-}^{j_3}(z^{(3)})
\end{array}
\Biggr) \notag \\
&\equiv \Biggl(~\psi_{\mathcal{P}}^J(z^{(1)},z^{(2)},z^{(3)})~\Biggr),
\end{align}
where $\mathcal{P}\equiv (\pm,\pm,\pm), (\pm,\pm,\mp),(\pm,\mp,\pm), (\mp,\pm,\pm)$ and $J\equiv (j_1,j_2,j_3)$.
In the same reason as $T^2$, depending on $M^{(1)}\lessgtr 0$, $M^{(2)}\lessgtr 0$ and $M^{(3)}\lessgtr 0$, only one of eight fields of $\psi_{\mathcal{P}}$ is well-defined, while the others cannot be normalizable.

Furthermore, the zero-mode fermions $\psi_{\mathcal{P}}^J(z^{(1)},z^{(2)},z^{(2)})$ are also constructed of $\psi_{\pm}^{j_g}(z^{(g)})$ on each torus, and from eqs.(\ref{psi_p_func}) and (\ref{psi_m_func}), $\psi_{\pm}^{j_g} (z^{(g)})$ satisfy eq.(\ref{psiZ_Ntra2}). 
Then, since $\psi_{\mathcal{P}}^J(z^{(1)},z^{(2)},z^{(3)})$ do not, in general, satisfy the physical state conditions (\ref{phys_st3_1}) and (\ref{phys_st3_2}) on $(T^2\times T^2\times T^2)/(Z_N\times Z_{N'})$, we may need to take appropriate linear combinations of them in order to obtain the physical states $\Psi_{\mathcal{P}}^J(z^{(1)},z^{(2)},z^{(3)})$.
Thus, the physical states are given by 
\begin{align}
\Psi_{\mathcal{P}}^J(z^{(1)},z^{(2)},z^{(3)}) 
&=\mathcal{N}'\sum_{\ell =0}^{N-1}\prod_{g=1,2}e^{i\pi \ell m_g(2j_g-(N-\ell)n_gt_g)/N} \notag \\
&\hspace{6mm} \times \sum_{\ell' =0}^{N'-1}\prod_{g'=2,3}e^{i\pi \ell' m'_{g'}(2j_{g'}-(N'-\ell')n'_{g'}t'_{g'})/N'} \notag \\
&\hspace{30mm}\times \psi_{\mathcal{P}}^{J+\ell T_{n_1n_2}+\ell' T' \hspace{-0.9mm} {}_{n'_2n'_3}}(z^{(1)},z^{(2)},z^{(3)}),
\end{align}
where $\mathcal{N}'$ is the normalization factor, $T_{n_1n_2}\equiv (n_1t_1,n_2t_2,0)$ and $T' \hspace{-0.9mm} {}_{n'_2n'_3}\equiv (0,n'_2t'_2,n'_3t'_3)$.

As a conclusion similar to $(T^2\times T^2)/Z_N$, we would like to notice that the number of zero-mode fermions for $(T^2\times T^2\times T^2)/(Z_N\times Z_{N'})$ is given by a multiple of $N$ and $N'$.
This result leads to an important conclusion that there is only one possibility to derive the three generations of matter, i.e.  $(N,N';M^{(1)},M^{(2)},M^{(3)})=(3,3;3,3,3)$ on $(T^2\times T^2 \times T^2)/(Z_N\times Z_{N'})$.
Furthermore, we consider the case of $(T^2\times T^2\times T^2)/(Z_N\times Z_{N'}\times Z_{N''})$, which also imposes the additional $Z_{N''}$ shift symmetry on the first and third tori, which relates the first torus with the third one as $(z^{(1)},z^{(2)},z^{(3)})\sim (z^{(1)}+e_{N''}^{m''_1n''_1},z^{(2)},z^{(3)}+e_{N''}^{m''_3n''_3})$.
Then, we obtain an important conclusion that there are only two possibilities to derive the three generations of matter, i.e. $(N,N',N'';M^{(1)},M^{(2)},M^{(3)})=(3,9,3;3,9,9),~(3,9,9;9,9,9)$, up to the permutation of parameters for the magnitude of fluxes and the shift symmetries. (See Appendix \ref{digNNN}.)
These results coincide with that of the operator formalism.

In the same way, the wave functions with $Z_N$ and $Z_{N'}$ charges, $\kappa$ and $\kappa'$, are given by
\begin{align}
\Psi_{\mathcal{P}\kappa \kappa'}^J(z^{(1)},z^{(2)},z^{(3)}) 
&=\mathcal{N}'\sum_{\ell =0}^{N-1}\prod_{g=1,2}e^{i\pi \ell m_g(2j_g-(N-\ell)n_gt_g)/N}e^{-2\pi i \ell \kappa /N}\notag \\
&\hspace{6mm}\times \sum_{\ell' =0}^{N'-1}\prod_{{g'}=2,3}e^{i\pi \ell' m'_{g'}(2j_b-(N'-\ell')n'_{g'}t'_{g'})/N'}e^{-2\pi i \ell' \kappa' /N'} \notag \\
&\hspace{30mm}\times \psi_{\mathcal{P}}^{J+\ell T_{n_1n_2}+\ell' T'
  \hspace{-0.9mm} {}_{n'_2n'_3}}(z^{(1)},z^{(2)},z^{(3)}),
\end{align}
where each of $\kappa$ and $\kappa'$ is some integer.

\subsection{Flavor structure}

Here we study the flavor structure in 
shifted orbifold models with magnetic fluxes.
First we give a brief review on the torus models \cite{Abe:2009vi}.
As seen in subsection \ref{RevU1T2}, if  $M>0$, 
the number of  zero-modes 
is equal to $M$ on $T^2$, 
and those wave functions are written by $\psi^j_+$ 
($j=0,\cdots,M - 1$) in eq.(\ref{psi_p_func}).
Each mode has the $Z_M$ charge $j$, 
which corresponds to the quantized coordinate or momentum in 
terms of  $\hat {\tilde Y}$ and $\hat {\tilde P}$.
Such $Z_M$ transformation is represented 
on 
\begin{eqnarray}\label{eq:multiplet}
\left( \begin{array}{c}
 \\
\psi^j  \\
         \\
\end{array}
\right) = 
\left(
\begin{array}{c}
\psi^0 \\
\psi^1 \\
\vdots \\
\psi^{M-1}
\end{array}
\right),
\end{eqnarray}
by 
\begin{eqnarray}\label{eq:Zg}
Z = \left(
\begin{array}{ccccc}
1 & & & & \\
  & \rho & & & \\
  & & \rho^2 & & \\
  & &   & \ddots & \\
  &  &  &    & \rho^{M-1} 
\end{array}
\right),
\end{eqnarray}
where $\rho = e^{2\pi i /M}$.
They also have another symmetry under the cyclic permutation, 
\begin{equation}\label{eq:permutation}
\psi^{j} \rightarrow \psi^{j+1} ,
\end{equation}
where $\psi^M=\psi^0$.
This cyclic permutation $Z_M^{(C)}$ is represented on the multiplet 
(\ref{eq:multiplet}) by
\begin{eqnarray}\label{eq:C}
C = \left(
\begin{array}{ccccc}
0 & 1& 0 &   \cdots & 0 \\
0  & 0 &1 &  \cdots & 0\\
&& & \ddots & \\
0 &0&0&\cdots & 1\\
1  &  0  & 0  & \cdots   & 0 
\end{array}
\right).
\end{eqnarray}
These generators, $Z$ and $C$, are not commutable, but 
satisfy the following algebraic relation,\footnote
{The symmetry, which is called the magnetic translational group, 
has been discussed in Ref.\cite{Sakamoto:2003, Tanimura:2002sf}.}
\begin{eqnarray}
CZ = \rho ZC.
\end{eqnarray}
Hence, the flavor symmetry including $Z$ and $C$ is a
non-Abelian symmetry.
Its diagonal elements are 
written by $Z^m (Z')^n$ with $m,n=0,\cdots, M-1$, 
where 
\begin{eqnarray}
Z' = \left(
\begin{array}{ccc}
\rho & &  \\
    & \ddots & \\
    &    & \rho 
\end{array}
\right),
\end{eqnarray}
on the multiplet (\ref{eq:multiplet}).
Then, this flavor symmetry would correspond to 
$(Z_M \times Z'_M) \rtimes Z_M^{(C)}$ on $T^2$ (see for review 
on non-Abelian discrete flavor symmetries \cite{Ishimori:2010au}).\footnote{
Similar flavor symmetries are obtained, e.g.
within the framework of heterotic orbifold models \cite{Kobayashi:2004ya}.}

Suppose that each of three tori has 
the magnetic flux corresponding to  $M^{(g)}$ ($g=1,2,3$), where 
$M^{(g)} > 0$.
Then, there are $M^{(g)}$ zero-modes on the $g$-th torus.
Their symmetry is the direct product of 
$\prod_{g=1}^3 (Z_{M^{(g)}} \times Z'_{M^{(g)}}) \rtimes Z_{M^{(g)}}^{(C)}$.
In order to realize the three-generation models, 
we choose the magnetic fluxes as 
$(M^{(1)},M^{(2)},M^{(3)})=(3,1,1)$ or their permutations.
In this model, only one torus, e.g. the first torus for 
$(M^{(1)},M^{(2)},M^{(3)})=(3,1,1)$ is important to the flavor structure.
That is, the three generations of fermions are quasi-localized 
at places different from each other on the first torus, 
while those sit at the same places on the other tori.
Thus, the zero-mode profiles on the first torus are 
important to realize the mass ratios between three generations, 
while the zero-mode profiles on the other tori 
are relevant to the overall strength 
of Yukawa couplings.
This model has the flavor symmetry 
$(Z_3 \times Z'_3) \rtimes Z_3^{(C)}$ isomorphic to 
$\Delta(27)$.

It would be obvious that the $T^2/Z_N$ model has 
a flavor structure similar to the above.
However, the $(T^2\times T^2)/Z_N$ model 
as well as the $(T^2\times T^2)/Z_N \times T^2$ model 
has a different flavor structure.
As an illustrating model, 
we consider the $(T^2\times T^2)/Z_3$ model, which leads to 
the three generations by choosing $M^{(1)}=M^{(2)}=3$. 
Before the $Z_3$ shifted orbifolding, 
there appear the three zero modes $\psi^{j_g}(z^{(g)})$ with 
$g=1,2$ and $j_g=0,1,2$ on each torus, and totally 
nine zero modes. 
They have the flavor symmetry 
$(Z_3 \times Z'_3) \rtimes Z_3^{(C)}$ on each torus and 
the total symmetry is their direct product.

By $Z_3$ shift orbifolding with $(m_1,n_1,m_2,n_2)=(1,0,1,0)$ 
corresponding to eq.(\ref{eq-phys-state}), only the three zero-modes 
$\psi^{j}(z^{(1)}) \otimes \psi^{3-j}(z^{(2)}) $ with $j=0,1,2$ 
remain, but the others $\psi^{j_1}(z^{(1)}) \otimes \psi^{j_2}(z^{(2)}) $ 
with $j_1 + j_2 \neq 0$ (mod 3) are projected out.
Through orbifolding, the $Z_{M^{(1)}=3}$ and $Z_{M^{(2)}=3}$ symmetries 
(\ref{eq:Zg}) on the first and second tori, respectively, 
are broken into the diagonal $Z_{3}$ one.
The other symmetries such as $Z'_{M^{(g)}}$ and $Z_{M^{(g)}}^{(C)}$ 
are also broken into the diagonal ones.
Then, totally the flavor symmetry $\Delta(27) \times \Delta(27)$  
is broken into the diagonal one $\Delta(27)$.
The flavor symmetry itself is the same as one in 
the three-generation model on $T^2\times T^2 \times T^2$ 
without shifted orbifolding.
However, in the three-generation model on $(T^2 \times T^2)/Z_3$, 
the zero-mode profiles of the three generations 
are localized at places different from each other 
on both the first and second tori.
That is, both tori are relevant to the flavor structure 
and mass ratios depend on geometrical aspects of 
both tori such as complex structure moduli.

We have studied quite simple models so far.
Furthermore, the flavor structure of shifted orbifold models 
can become richer in slightly extended models.
Suppose that there is an additional $U(1)$ gauge symmetry.
We do not introduce the magnetic flux background 
for the additional $U(1)$, but we embed $Z_3$ shift orbifolding 
into this additional $U(1)$.
That is, the fermion with the additional $U(1)$ charge $q'$, 
which is integer, has the phase $e^{2\pi i q'/3}$.
In this case, the zero-modes, 
$\psi^{j}(z_1) \otimes \psi^{3-j}(z_2) $ with $j=0,1,2$, 
do not survive in the above model, 
but the zero-modes  $\psi^{j}(z_1) \otimes \psi^{3-j+k}(z_2) $ with
$j=0,1,2$ and $k=-q'$, survive through the $Z_3$ shift orbifolding.
The surviving number, i.e. 3, does not change, but 
the combinations of surviving wave functions depend on the $U(1)$ charge
$q'$.
Hence, the flavor structure becomes rich.
For example, when this charge $q'$ corresponds 
to the hypercharge,  the three generations of quarks and leptons 
would have quite interesting flavor structure.
We will study such model building and its flavor structure elsewhere.

%%%%%%%%%%%%%%%%%%%%%%%%%%%%%%%%%%%%%%%%%%%%%%%%%%%%%%%%%%%%%%%%%%%%%%%%%%%%%%%
\section{Conclusions}

We have studied the $U(1)$ gauge theory on some shifted orbifolds with magnetic flux and proposed a mechanism to obtain the generation of matter in the standard model.
On the space, we consider the behavior of fermions in two different means.
One is the operator formalism for the quantum mechanical system and the other is the wave functions for the field theory.
The operator formalism turns out to be useful to analyze the general structure of the spectrum.
On the other hand, the wave function approach becomes important on computing the 4D Yukawa coupling of phenomenological models on the shifted orbifolds that we consider in this paper.
We investigated the relations between the magnetic fluxes and the number of degeneracy of zero-mode fermions in both approaches and showed the results to be consistent with each other.

Then we found that the number of degeneracy of zero-mode fermions is related to $N$ of $Z_N$, that is, the geometry of space such as $(T^2\times T^2)/Z_N$.
Actually, while there existed no constraint for the degeneracy of zero-mode fermions on $T^2/Z_N$, we obtained the constraint on the degeneracy of zero-mode fermions on $(T^2\times T^2)/Z_N$, that is to say, {\it the number of degeneracy of zero-mode fermions is always a multiple of} $N$.
This result is phenomenologically very important, because we have a unique choice of $(N;M_1,M_2)=(3;3,3)$ if we want to construct a model deriving the three generations of matter on $M^4\times (T^2\times T^2)/Z_N$ with the magnetic fluxes $(M^{(1)},M^{(2)})$.
In a similar way, we found some candidates for the models to derive the three generations of matter in cases of $(T^2 \times T^2 \times T^2)/(Z_N \times Z_{N'})$ and $(T^2 \times T^2 \times T^2)/(Z_N \times Z_{N'} \times Z_{N''})$.
In the case of $(T^2 \times T^2 \times T^2)/(Z_N \times Z_{N'})$, the candidate to derive the three generations of matter is that $(N,N';M^{(1)},M^{(2)},M^{(3)})=(3,3;3,3,3)$.
In the case of $(T^2\times T^2\times T^2)/(Z_N\times Z_{N'}\times Z_{N''})$, the candidates to derive the three generations of matter are that $(N,N',N'';M^{(1)},M^{(2)},M^{(3)})=(3,9,3;3,9,9),~(3,9,9;9,9,9)$ up to the permutation of parameters for the magnitude of fluxes and the shift symmetries.
Thus, we may conclude that a very restricted class of shifted orbifold models can produce the three generations of matter, in general.

We comment on the difference between the shifted orbifold and 
the twisted orbifold.
On the twisted orbifold, there are fixed points.
Thus, there is the degree of freedom to put localized matter fields 
on the fixed points in the twisted orbifold models 
with magnetic fluxes.
There is no such a degree of freedom in the shifted 
orbifold models, because there is no fixed points.
Hence, the spectrum of the shifted orbifold models 
is completely determined by the shift and magnetic fluxes.

For the three-generation models, 
the torus models without shifted orbifolding and 
the shifted orbifold models would lead to 
the same flavor symmetry, i.e. $\Delta(27)$.
However, while only the one of tori is relevant to 
the flavor structure in the former, 
two or three tori are important to the flavor structure 
in the latter.
These behaviors would lead to phenomenologically 
interesting aspects.
We would study realistic model building 
and its phenomenological aspects elsewhere.

%%%%%%%%%%%%%%%%%%%%%%%%%%%%%%%%%%%%%%%%%%%%%%%%%%%%%%%%%%%%%%%%%%%%%%%%%%%%%%%%%%%%%
\section*{Acknowledgements}
This work was supported in part by scientific grants from the Ministry of Education, Culture,
Sports, Science and Technology under Grant No.~24340049 (T.K.), No.~23$\cdot$9368 (T.M.), No.~22540281 and No.~20540274 (M.S.).
K.N. is partially supported by funding available from the
Department of Atomic Energy, Government of India for the Regional
Centre for Accelerator-based Particle Physics (RECAPP), Harish-Chandra
Research Institute.

%\pagebreak
%%%%%%%%%%%%%%%%%%%%%%%%%%%%%%%%%%%%%%%%%%%%%%%%%%%%%%%%%%%%%%%%%%%%%%%%%%%%%%%
\appendix
\section{$Z_{N}$ shifted orbifolding and basis transformation\label{soabt}}
In this appendix, we discuss the general form of $Z_{N}$ shifted orbifold and its transformation into the simple form using a basis transformation of the torus bases. we can define the $T^2/Z_N$ shifted orbifold as the general identification,
	%%%%%%%%%%%%%%%%%%%%%%%%%%%%%%%%%%%%%%%%%%%%%%
	\begin{align}
	Z_{N}:{\bm y} \sim {\bm y}+\frac{1}{N}(r_{1}{\bm u}_{1}+r_{2}{\bm u}_{2}),\label{ZN}
	\end{align}
	%%%%%%%%%%%%%%%%%%%%%%%%%%%%%%%%%%%%%%%%%%%%%%
where $N$ is some positive integer, and the g.c.d. of the integers $r_{1}$ and $r_{2}$, say $r$, is relatively prime with $N$.  However, a choice of lattice bases $({\bm u}_{1}, {\bm u}_{2})$ is not unique and we can take another lattice bases $({\bm u'}_{1}, {\bm u'}_{2})$ using a matrix $U\in SL(2,\mathbb{Z})$  as
	%%%%%%%%%%%%%%%%%%%%%%%%%%%%%%%%%%%%%%%%%%%%%%
	\begin{align}
	\left(\begin{array}{l}
		{\bm u'}_{1}\\
		{\bm u'}_{2} 
		\end{array}\right) 
		= U 
		\left(\begin{array}{l}
		{\bm u}_{1}\\
		{\bm u}_{2} 
		\end{array}\right),
	\end{align}
	%%%%%%%%%%%%%%%%%%%%%%%%%%%%%%%%%%%%%%%%%%%%%%
and a suitable choice of a new basis leads to the simple shifted orbifold form like eq.(\ref{simpleZN}). Since we assumed that $r$ is the g.c.d. of $r_{1}$ and $r_{2}$, both of them can be expressed as 
	%%%%%%%%%%%%%%%%%%%%%%%%%%%%%%%%%%%%%%%%%%%%%%
	\begin{align}
	r_{1}=\alpha r,~~r_{2}=\beta r, 
	\end{align}
	%%%%%%%%%%%%%%%%%%%%%%%%%%%%%%%%%%%%%%%%%%%%%%
where each of $\alpha$ and $\beta$ is some integer.
Since $\alpha$ and $\beta$ are relatively prime with each other, there exist some integers $\gamma$ and $\delta$ such that	
	%%%%%%%%%%%%%%%%%%%%%%%%%%%%%%%%%%%%%%%%%%%%%%
	\begin{align}
	\alpha \delta -\beta \gamma =1~~~ \mathrm{for}~~{}^{\exists}\gamma, {}^{\exists}\delta \in \mathbb{Z}~.
	\end{align}
	%%%%%%%%%%%%%%%%%%%%%%%%%%%%%%%%%%%%%%%%%%%%%%
Constructing the $SL(2,\mathbb{Z})$ matrix
	%%%%%%%%%%%%%%%%%%%%%%%%%%%%%%%%%%%%%%%%%%%%%%
	\begin{align}
	U=\left( \begin{array}{cc}
			\alpha &\beta \\
			\gamma &\delta
		     \end{array}
		     \right),
	\end{align}
	%%%%%%%%%%%%%%%%%%%%%%%%%%%%%%%%%%%%%%%%%%%%%%
we can take a new bases 
	%%%%%%%%%%%%%%%%%%%%%%%%%%%%%%%%%%%%%%%%%%%%%%
	\begin{align}
	\left(\begin{array}{l}
		{\bm u'}_{1}\\
		{\bm u'}_{2} 
		\end{array}\right) 
		= U
		\left(\begin{array}{l}
		{\bm u}_{1}\\
		{\bm u}_{2} 
		\end{array}\right),
	\end{align}
	%%%%%%%%%%%%%%%%%%%%%%%%%%%%%%%%%%%%%%%%%%%%%%
in which the shifted orbifold (\ref{ZN}) is written as
	%%%%%%%%%%%%%%%%%%%%%%%%%%%%%%%%%%%%%%%%%%%%%%
	\begin{align}
	Z_{N}:{\bm y}\sim {\bm y}+\frac{r}{N}{\bm u}_{1}'.
	\label{T2ZN_ZNtra}
	\end{align}
	%%%%%%%%%%%%%%%%%%%%%%%%%%%%%%%%%%%%%%%%%%%%%%
Since $r$ and $N$ are relatively prime with each other, they satisfy
	%%%%%%%%%%%%%%%%%%%%%%%%%%%%%%%%%%%%%%%%%%%%%%
	\begin{align}
	pr -qN=1~~~ \mathrm{for}~~{}^{\exists}p,{}^{\exists}q \in \mathbb{Z}~.
	\end{align}
	%%%%%%%%%%%%%%%%%%%%%%%%%%%%%%%%%%%%%%%%%%%%%%
Using the above integer $p$, we can define the new identification as
 	%%%%%%%%%%%%%%%%%%%%%%%%%%%%%%%%%%%%%%%%%%%%%%
	\begin{align}
	Z'_{N}: {\bm y}\sim {\bm y}+\frac{pr}{N}{\bm u}_{1}\sim {\bm y}+\frac{1}{N}{\bm u}_{1}, 
	\end{align}
	%%%%%%%%%%%%%%%%%%%%%%%%%%%%%%%%%%%%%%%%%%%%%%
up to the torus identification. 

In a similar way, we can define the $(T^2\times T^2)/Z_N$ shifted orbifold as the general identification
	%%%%%%%%%%%%%%%%%%%%%%%%%%%%%%%%%%%%%%%%%%%%%%
	\begin{align}
	Z_{N}:({\bm y}^{(1)},{\bm y}^{(2)}) \sim \Bigl({\bm y}^{(1)}+\frac{1}{N}(r_{11}{\bm u}^{(1)}_{1}+r_{12}{\bm u}^{(1)}_{2}),{\bm y}^{(2)}+\frac{1}{N}(r_{21}{\bm u}^{(2)}_{1}+r_{22}{\bm u}^{(2)}_{2})\Bigr), \notag \\
	\end{align}
	%%%%%%%%%%%%%%%%%%%%%%%%%%%%%%%%%%%%%%%%%%%%%%
where $N$ is some positive integer, each of $r_{gj}~(g,j=1,2)$ is some integer and $r^{(g)}$ which are defined as the g.c.d. of $r_{g1}$ and $r_{g2}$ are relatively prime with $N$.   
Changing the lattice bases like eq.(\ref{T2ZN_ZNtra}), we can rewrite this identification as
	%%%%%%%%%%%%%%%%%%%%%%%%%%%%%%%%%%%%%%%%%%%%%%
	\begin{align}
	Z_{N}:({\bm y}^{(1)},{\bm y}^{(2)}) \sim \Bigl({\bm y}^{(1)}+\frac{r^{(1)}}{N}{\bm u}^{(1)}_{1},{\bm y}^{(2)}+\frac{r^{(2)}}{N}{\bm u}^{(2)}_{1}\Bigr).
	\end{align}
	%%%%%%%%%%%%%%%%%%%%%%%%%%%%%%%%%%%%%%%%%%%%%%
We have to note that each of $r^{(g)}$ is relatively prime with $N$, i.e.
	%%%%%%%%%%%%%%%%%%%%%%%%%%%%%%%%%%%%%%%%%%%%%%
	\begin{align}
	p_gr^{(g)} -q_gN=1~~~ \mathrm{for}~~{}^{\exists}p_g,{}^{\exists}q_g \in\mathbb{Z}~.
	\end{align}
	%%%%%%%%%%%%%%%%%%%%%%%%%%%%%%%%%%%%%%%%%%%%%%
Using $p_2$, we can put $r^{(2)}$ to 1 and define the new identification as 
	%%%%%%%%%%%%%%%%%%%%%%%%%%%%%%%%%%%%%%%%%%%%%%
	\begin{align}
	Z'_{N}:({\bm y}^{(1)},{\bm y}^{(2)}) \sim \Bigl({\bm y}^{(1)}+\frac{d}{N}{\bm u}^{(1)}_{1},{\bm y}^{(2)}+\frac{1}{N}{\bm u}^{(2)}_{1}\Bigr),
	\end{align}
	%%%%%%%%%%%%%%%%%%%%%%%%%%%%%%%%%%%%%%%%%%%%%%
where 
	%%%%%%%%%%%%%%%%%%%%%%%%%%%%%%%%%%%%%%%%%%%%%%
	\begin{align}
	d\equiv p_2 r^{(1)}~>0.
	\end{align}
	%%%%%%%%%%%%%%%%%%%%%%%%%%%%%%%%%%%%%%%%%%%%%%
The above $Z'_{N}$ shifted orbifolding is nothing but eq.(\ref{T2T2Znidentification}).

In the case of $(T^2\times T^2\times T^2)/(Z_{N}\times Z_{N'})$, the situation is a little bit different. For the $Z_{N}$ shifted orbifold, we can apply the same argument which leads us to the simple shifted orbifold
	%%%%%%%%%%%%%%%%%%%%%%%%%%%%%%%%%%%%%%%%%%%%%%
	\begin{align}
	Z_{N}:({\bm y}^{(1)},{\bm y}^{(2)},{\bm y}^{(3)}) \sim \Bigl({\bm y}^{(1)}+\frac{d}{N}{\bm u}^{(1)}_{1},{\bm y}^{(2)}+\frac{1}{N}{\bm u}^{(2)}_{1},{\bm y}^{(3)}\Bigr), 
	\end{align}
	%%%%%%%%%%%%%%%%%%%%%%%%%%%%%%%%%%%%%%%%%%%%%%
where each of $N$ and  $d$ is some positive integer and $d$ is relatively prime with $N$.
However, for the $Z_{N'}$ shifted orbifold, we cannot apply the same
argument at the same time since we are unable to execute  a basis transformation for the second torus anymore. It is  because that we already executed a basis transformation for the second torus to simplify the $Z_{N}$ shifted orbifold. Thus, for the second torus, we have to assume the general form when we consider the $Z_{N'}$ shifted orbifold. However, for the third torus, we can still execute a basis transformation which means that $Z_{N'}$ shifted orbifold is given by
	%%%%%%%%%%%%%%%%%%%%%%%%%%%%%%%%%%%%%%%%%%%%%%
	\begin{align}
	Z_{N'}:({\bm y}^{(1)},{\bm y}^{(2)},{\bm y}^{(3)}) \sim \Bigl({\bm y}^{(1)},{\bm y}^{(2)}+\frac{1}{N'}(s_{1}{\bm u}^{(2)}_{1}+s_{2}{\bm u}^{(2)}_{2}),{\bm y}^{(3)}+\frac{d'}{N'}{\bm u}^{(3)}_{1}\Bigr),
	\end{align}
	%%%%%%%%%%%%%%%%%%%%%%%%%%%%%%%%%%%%%%%%%%%%%%
where each of $N'$ and $d'$ is some positive integer and each of $s_{1}$ and $s_{2}$ is some integer.
When we define $s'$ as the g.c.d. of $s_{1}$ and $s_{2}$, the g.c.d. of $s'$ and $d'$ is relatively prime with $N'$.

\section{The degeneracy of spectrum on $(T^2\times T^2\times T^2)/(Z_{N}\times Z_{N'}\times Z_{N''})$\label{digNNN}}

We here discuss the degeneracy of the spectrum on $(T^2\times T^2\times
T^2)/(Z_{N}\times Z_{N'}\times Z_{N''})$ shifted orbifold with the identifications (\ref{T2T2T2/ZNZNZN_iden}).
We define $N$, $N'$ and $N''$ as
%%%%%%%%%%%%%
\begin{align}
N\equiv l_1d_{12}d_{13}d_{123},~~N'\equiv l_2d_{12}d_{23}d_{123},~~N''\equiv
l_3d_{13}d_{23}d_{123},
\end{align}
%%%%%%%%%%%%%
where any pair of $l_1$, $l_2$ and $l_3$ are relatively prime with each
other, $d_{123}$ is the g.c.d. of $N$, $N'$ and $N''$, $d_{12}$ is
relatively prime with each of $l_1$, $l_2$ and $d_{123}$, $d_{23}$ is
relatively prime with each of $l_2$, $l_3$ and $d_{123}$, and $d_{13}$ is
relatively prime with each of $l_1$, $l_3$ and $d_{123}$.
Since $d_1$ is the g.c.d. of $N$ and $N'$, $d_2$ is the g.c.d. of $N'$ and
$N''$ and $d_3$ is the g.c.d. of $N''$ and $N$, we can rewrite $d_1$, $d_2$
and $d_3$ as $d_1=d_{13}d_{123}$, $d_2=d_{12}d_{123}$ and
$d_3=d_{23}d_{123}$, respectively.
The magnitude of flux on each torus turns out to be of the form
%%%%%%%%%%%%%
\begin{align}
M^{(1)}=t_1l_1l_3d_{13}d_{12}d_{23}d_{123},~~M^{(2)}=t_2l_1l_2d_{13}d_{12}d_
{23}d_{123},~~M^{(3)}=t_3l_2l_3d_{13}d_{12}d_{23}d_{123},\notag \\
\end{align}
%%%%%%%%%%%%%
where each of $t_{1}$, $t_{2}$ and $t_{3}$ is some integer.
Then it follows that the number of degeneracy is given by
%%%%%%%%%%%%%
\begin{align}
\frac{|M^{(1)}M^{(2)}M^{(3)}|}{NN'N''}
  = |t_1t_2t_3| \frac{NN'N''}{d_1d_2d_3},
\end{align}
%%%%%%%%%%%%%
which is the result of eq.(\ref{deg_T6NNN}).
When we want to construct $(T^2\times T^2\times T^2)/(Z_{N}\times
Z_{N'}\times Z_{N''})$ shifted orbifold models with the three generations,
there are only two possibilities up to the permutation of parameters for the
magnitude of fluxes and the shift symmetries.
One is $(N,N',N'';M^{(1)},M^{(2)},M^{(3)})=(3,9,3;3,9,9)$ on
$t_1=t_2=t_3=1$, $l_1=l_3=d_{12}=d_{13}=d_{23}=1$ and $l_2=d_{123}=3$.
The other is $(N,N',N'';M^{(1)},M^{(2)},M^{(3)})=(3,9,9;9,9,9)$ on
$t_1=t_2=t_3=1$, $l_1=l_2=l_3=d_{12}=d_{13}=1$ and $d_{23}=d_{123}=3$.

\section{Redefinition of fields and $\alpha_{i}$ parameters\label{Wlp&alp}}

We consider the relation between the redefinition of fields $A(z,\bar{z})$ and $\Phi(z,\bar{z})$ and $\alpha_{i}(\in \mathbb{R})$ parameters in $\chi_{i}(z,\bar{z})$, which are given by 
\begin{align}
\chi_1(z,\bar{z})={b\over 2\mathrm{Im}\tau}\mathrm{Im}(z+a)+{\pi \alpha_1\over q},~~\chi_2(z,\bar{z})={b\over 2\mathrm{Im}\tau}\mathrm{Im}(\bar{\tau}(z+a))+{\pi \alpha_2\over q}.
\end{align}
Let us redefine $\Phi(z,\bar{z})$ in eq.(\ref{Phi_T^2}), which has the $U(1)$ charge $q$, by
\begin{align}
\Phi(z,\bar{z})\equiv e^{iq\mathrm{Re}(\bar{\gamma} z)}\tilde{\Phi}(z,\bar{z}),
\end{align}
where $\gamma$ is any complex number.
With this the redefinition, the covariant derivatives for $\Phi$ can be written by
\begin{align}
&(\partial_z -iqA_z)\Phi(z,\bar{z})=e^{iq\mathrm{Re}(\bar{\gamma} z)}(\partial_z -iq\tilde{A}_z)\tilde{\Phi}(z,\bar{z}), \notag \\
&(\partial_{\bar{z}} -iqA_{\bar{z}})\Phi(z,\bar{z})=e^{iq\mathrm{Re}(\bar{\gamma} z)}(\partial_{\bar{z}} -iq\tilde{A}_{\bar{z}})\tilde{\Phi}(z,\bar{z}).
\end{align}
Here we defined $\tilde{A}_z$ and $\tilde{A}_{\bar{z}}$ as $\tilde{A}_z\equiv A_z-\bar{\gamma}/2$ and $\tilde{A}_{\bar{z}}\equiv A_{\bar{z}}-\gamma/2$, respectively.
Then the Wilson line phases of $\tilde{A}$ are given by $\tilde{a}\equiv a-\gamma\mathrm{Im}\tau /b$.

We notice that under the transformation $\Phi \to \tilde{\Phi}$ and $A \to \tilde{A}$,
the Lagrangian density $\mathcal{L}$ is invariant, i.e. $\mathcal{L}(A,\Phi)=\mathcal{L}(\tilde{A},\tilde{\Phi})$.
Defining $\tilde{\chi}_i$ as
\begin{align}
\tilde{\chi}_1(z,\bar{z})\equiv \chi_1(z,\bar{z})-\mathrm{Re}\gamma,~~\tilde{\chi}_2(z,\bar{z})\equiv \chi_2(z,\bar{z})-\mathrm{Re}(\bar{\tau} \gamma),
\end{align}
we can check that $\tilde{A}$ and $\tilde{\Phi}$ with $\tilde{\chi}_i$ satisfy  
\begin{align}
&\tilde{A}(z+1,\bar{z}+1)=\tilde{A}(z,\bar{z})+d\tilde{\chi}_1(z,\bar{z}),~~\tilde{A}(z+\tau,\bar{z}+\bar{\tau})=\tilde{A}(z,\bar{z})+d\tilde{\chi}_2(z,\bar{z}),\notag \\
&\tilde{\Phi}(z+1,\bar{z}+1)=e^{iq\tilde{\chi}_1(z,\bar{z})}\tilde{\Phi}(z,\bar{z}),~~\tilde{\Phi}(z+\tau,\bar{z}+\bar{\tau})=e^{iq\tilde{\chi}_2(z,\bar{z})}\tilde{\Phi}(z,\bar{z}).
\end{align}
If we take $\gamma$ to satisfy $\pi \alpha_1 -q\mathrm{Re}\gamma =0$ and $\pi \alpha_2 -q\mathrm{Re}(\bar{\tau} \gamma) =0$, we obtain $\tilde{a}=a-i\pi (\alpha_1\bar{\tau}+\alpha_2)/qb$.
Thus, since we can take any $\gamma$, we can always make $\alpha_i$ absorbed into the Wilson line phase $a$ by the redefinition of fields.
This result can be applied in the multi-torus case.

%%%%%%%%%%%%%%%%%%%%%%%%%%%%%%%%%%%%%%%%%%%%%%%%%%%%%%%%%%%%%%%%%%%%%%%%%%%%%%%%%%%%%

\end{document}